\documentclass[12pt,a4paper]{article}

\usepackage[T1]{fontenc}
\usepackage[utf8]{inputenc}
\usepackage{lmodern}
\usepackage{amsmath}
\usepackage{amssymb}
\usepackage{graphicx}
\usepackage{caption}
\usepackage{envmath}
\usepackage{cancel}
\usepackage{braket}
\usepackage{slashed}
\usepackage{jheppub}

\newcommand{\mrm}[1]{_{\rm #1}}

\begin{document}

\begin{flushright}
CERN-TH-2018-146
\end{flushright}

\vspace{1cm}

\begin{center}
 \large\textbf{AlterBBN v2: A public code for calculating Big-Bang nucleosynthesis constraints in alternative cosmologies}
\end{center}

\vspace{1cm}

\begin{center}
A. Arbey$^{a,b,}$\footnote{\texttt{alexandre.arbey@ens-lyon.fr}}$^,$\footnote{Also Institut Universitaire de France, 103 boulevard Saint-Michel, 75005 Paris, France}, J. Auffinger$^{a,c,}$\footnote{\texttt{jeremy.auffinger@ens-lyon.fr}}, K. P. Hickerson$^{d,}$\footnote{\texttt{kevinh@caltech.edu}}, E. S. Jenssen$^{e,}$\footnote{\texttt{EspenJenssen@hotmail.com}}\\[0.4cm]
{\sl$^a$Univ Lyon, Univ Lyon 1, CNRS/IN2P3, Institut de Physique Nucl\'eaire de Lyon, UMR5822, F-69622 Villeurbanne, France}\\
\vspace{0.2cm}
{\sl$^b$Theoretical Physics Department, CERN, CH-1211 Geneva 23, Switzerland}\\
\vspace{0.2cm}
{\sl$^c$Ecole Normale Sup\'erieure de Lyon, F-69342 Lyon, France}\\
\vspace{0.2cm}
{\sl$^d$W. K. Kellogg Radiation Laboratory, California Institute of Technology, Pasadena, California 91125, USA}\\
\vspace{0.2cm}
{\sl$^e$Institute of Theoretical Astrophysics, The University of Oslo, Boks 1072 Blindern, NO-0316 Oslo, Norway}
\end{center}

\vspace{1.cm}

URL: \url{https://alterbbn.hepforge.org/}

\vspace{1.cm}

\begin{center}
{\bf Abstract}
\end{center}

We present the version 2 of \texttt{AlterBBN}, an open public code for the calculation of the abundance of the elements from Big-Bang nucleosynthesis. It does not rely on any closed external library or program, aims at being user-friendly and allowing easy modifications, and provides a fast and reliable calculation of the Big-Bang nucleosynthesis constraints in the standard and alternative cosmologies.

\newpage

\tableofcontents{}

\newpage

\section{Introduction}

\texttt{AlterBBN} is a public C program for the calculation of the abundance of the elements generated during Big-Bang nucleosynthesis (BBN), released under the GPL version 3 license. The first version was released in 2011 \cite{2012CoPhC.183.1822A} and could be considered as a spiritual successor of \texttt{NUC123} \cite{Kawano1992}. Contrary to other public BBN codes such as the Fortran program \texttt{PArthENoPE} \cite{2008CoPhC.178..956P,2018CoPhC.233..237C} or the Mathematica code \texttt{PRIMAT} \cite{2018arXiv180108023P}, the main purpose of \texttt{AlterBBN} is to provide a fast and reliable calculation of the abundance of the elements in the standard model of cosmology as well as in alternative scenarios. \texttt{AlterBBN} is also included in the \texttt{SuperIso Relic} package \cite{2010CoPhC.181.1277A,2011CoPhC.182.1582A,2018arXiv180611489A}.

\texttt{AlterBBN} can be downloaded from its new website:
\begin{center}
	\texttt{https://alterbbn.hepforge.org/}
\end{center}
and involves an enlarged development team.

In \texttt{AlterBBN v2}, automatic calculation of errors and correlations has been implemented using methods similar to the ones described in Refs.~\cite{1998PhRvD..58f3506F,2016JHEP...11..097A}. To improve the speed of the calculations, parallel processing is possible through the \texttt{OpenMP} library. In addition, the units throughout the code have been unified to have GeV as the main unit, in order to stay consistent with \texttt{SuperIso Relic}. The nuclear reaction network has also been extended, and the code has been scrutinised and improved for precision and speed.

Moreover, new cosmological scenarios have been implemented in \texttt{AlterBBN}, such as reheating, decaying primordial scalar field, and WIMPs.

The rest of this paper is organised as follows. Section 2 provides a review of BBN physics and the cosmological modifications implemented in \texttt{AlterBBN}. Section 3 describes the content of the \texttt{AlterBBN} package. Section 4 gives usage instructions. Section 5 describes the input and output of \texttt{AlterBBN}. Section 6 provides examples of analyses which can be performed with \texttt{AlterBBN}. Short descriptions of the nuclear reaction network, integration methods and BBN constraints implemented in \texttt{AlterBBN} are given in the appendices.

\section{BBN physics and cosmology}

In this section we briefly present the physics relevant for BBN studies. We consider the system of natural units $c = \hbar = k\mrm{B} = 1$.

\subsection{Cosmological standard model and BBN}

\subsubsection{General equations}

At the beginning of the BBN epoch, the Universe contains photons $\gamma$, electrons $e^-$ and positrons $e^+$, protons $p$ and neutrons $n$, neutrinos $\nu$ and presumably dark matter $\chi$. During BBN, new nuclei will form over nuclear reactions (see Table \ref{table.nuclei} in Appendix \ref{appendix.nuclear}), which are contained in the general name of baryons b. BBN takes place in the more global frame of the early Universe expansion, parametrised by the expansion rate $\dot{a}$ given by the Friedmann equation as a function of the total density $\rho\mrm{tot}$:
\begin{equation}
	H^2 = \left( \frac{\dot{a}}{a} \right)^2 = \frac{8\pi G}{3}\rho\mrm{tot}\,, \label{eq_a}
\end{equation}
where $H$ is the Hubble parameter and $G$ is the Newton gravitational constant. The total density $\rho\mrm{tot}$ is given by the sum over all the aforementioned constituents:
\begin{equation}
	\rho\mrm{tot} = \rho_\gamma + \rho_\nu + \rho\mrm{b} + \rho_{e^-} + \rho_{e^+} + \rho_\chi\,.
\end{equation}
This is completed by the equations of continuity of each set of independent components:
\begin{equation}
	\frac{{d}}{{d}t}(\rho\mrm{set}a^3) + P\mrm{set}\frac{{d}}{{d}t}(a^3) - T \frac{{d}}{{d}t}(s\mrm{set} a^3)  = 0\,. \label{eq_rho}
\end{equation}
In the most general cases, both $\rho\mrm{set}$ and $s\mrm{set}$ can be considered as functions of both temperature $T$ and scale factor $a$, or -- since the scale factor is only time-dependent -- as explicit functions of both temperature $T$ and time $t$. Using the fact that 
\begin{equation}
	\frac{d\ln(a^3)}{dt} = 3 H \,,
\end{equation}
the continuity equation can be translated into a relation between the scale factor and temperature:
\begin{equation}
	\frac{{d\ln(a^3)}}{{d}t} = -3 H \; \frac{\dfrac{\partial \rho\mrm{set}}{\partial T} - T \dfrac{\partial s\mrm{set}}{\partial T}}{\dfrac{\partial \rho\mrm{set}}{\partial t} + 3 H (\rho\mrm{set} + P\mrm{set}) - T \left(\dfrac{\partial s\mrm{set}}{\partial t} + 3 H s\mrm{set}\right)}\,. \label{eq:dlna3_dt}
\end{equation}
The cosmological components are generally considered to be ideal gases, with by construction densities depending only on the temperature, and expansion is assumed to be adiabatic so that $d(s a^3)/dt = 0$ and $\partial s/\partial T = 0$, this equation simplifies into:
\begin{equation}
	\frac{{d\ln(a^3)}}{{d}t} = - \;\frac{\dfrac{d\rho\mrm{set}}{dT}}{\rho\mrm{set} + P\mrm{set}}\,. \label{eq:dlna3_dt_simp}
\end{equation}
In this equation, since dark matter and neutrinos can be considered in the standard case as decoupled, the density $\rho\mrm{set}$ and pressure $P\mrm{set}$ of the set of interacting components are given by:
\begin{equation}
	\rho\mrm{set} = \rho_\gamma + \rho\mrm{b} + \rho_{e^-} + \rho_{e^+}\,,
\end{equation}
\begin{equation}
	P\mrm{set} = P_\gamma + P\mrm{b} + P_{e^-} + P_{e^+}\,.
\end{equation}
They can be computed through statistical mechanics to give for photons:
\begin{equation}
	\begin{matrix}
		\rho_\gamma = \dfrac{\pi^2}{15}T^4\,, && && P_\gamma = \dfrac{1}{3}\rho_\gamma\,,
	\end{matrix}
\end{equation}
and for neutrinos:
\begin{equation}
	\begin{matrix}
		\rho_\nu = N_\nu \dfrac{7}{8}\dfrac{\pi^2}{15}T_\nu^4\,, && && P_\nu = \dfrac{1}{3}\rho_\nu\,, \label{eq_nu}
	\end{matrix}
\end{equation}
where $N_\nu$ is the number of Standard Model neutrino species, modified by the non-exact relativistic behaviour of $e^\pm$ to the value $N_\nu = 3.046$ \cite{2015PhRvD..91h3505N}. The temperature discrepancy $T_\nu/T = (4/11)^{1/3}$ comes from the neutrino decoupling which happens before BBN.

On the other hand, we parametrise the sums of $e^\pm$ densities and pressures thanks to the modified Bessel functions $K_i$ \cite{Chandra1939}:
\begin{equation}
	\rho_{e^-} + \rho_{e^+} = \frac{2}{\pi^2}m_e^4 \sum_{n=1}^{\infty} (-1)^{n+1}\cosh(n\phi_e)M(nz)\,, \label{eq_rho_epm}
\end{equation}
\begin{equation}
	P_{e^-} + P_{e^+} = \frac{2}{\pi^2}m_e^4 \sum_{n=1}^{\infty} \frac{(-1)^{n+1}}{nz}\cosh(n\phi_e)L(nz)\,, \label{eq_P_epm}
\end{equation}
where we have defined the adimensioned electron mass $z = m_e/T$ and chemical potential $\phi_{e^-} = -\phi_{e^+} \equiv \phi_e = \mu_e/T$ and:
\begin{equation}
	\begin{matrix}
		L(z) = \dfrac{K_2(z)}{z}\,, && && M(z) = \dfrac{1}{z}\left( \dfrac{3}{4}K_3(z) + \dfrac{1}{4}K_1(z) \right)\,.
	\end{matrix}
\end{equation}
For the purpose of our computation, these convergent sums will be truncated at $n=7$. Another equation comes from the charge conservation of the Universe. The difference between the $e^\pm$ densities is linked to the nuclei abundances through:
\begin{equation}
	n_{e^-} - n_{e^+} = \frac{h_\eta T^3 S}{M\mrm{u}}\,, \label{eq_ne1}
\end{equation}
with:
\begin{equation}
	S = \sum_i Z_i Y_i\,,
\end{equation}
where $Z_i$ and $Y_i$ are the charge number and abundance of nucleus $i$ respectively. The variable $h$ parameterises the baryon-to-photon ratio in the following way \cite{1969ApJS...18..247W,Kawano1992,1967ApJ...148....3W}:
\begin{equation}
	h_\eta(T) = M\mrm{u} \frac{n_\gamma(T)}{T^3}\eta(T)\,,
\end{equation}
where $M\mrm{u}$ is the unit atomic mass and $\eta$ the baryon-to-photon ratio. The difference in Eq. (\ref{eq_ne1}) can also be parametrised as:
\begin{equation}
	n_{e^-} - n_{e^+} = \frac{2}{\pi^2}m_e^3 \sum_{i=1}^{\infty}(-1)^{n+1}\sinh(n\phi_e)L(nz)\,. \label{eq_ne2}
\end{equation}
Using Eqs. (\ref{eq_ne1}) and (\ref{eq_ne2}) the electron chemical potential can be determined through:
\begin{equation}
	\frac{d\phi_e}{dt} = \frac{\partial \phi_e}{\partial T}\frac{dT}{dt} + \frac{\partial \phi_e}{\partial a}\frac{{d}a}{{d}t} + \frac{\partial\phi_e}{\partial S}\frac{{d}S}{{d}t}\,. \label{eq_phie}
\end{equation}

Finally, the baryon density and pressure are given by the sums on the $i$ nuclei~\cite{1969ApJS...18..247W}:
\begin{equation}
	\rho\mrm{b} = h_\eta T^3\left( 1+\sum_i\left( \dfrac{\Delta M_i}{M\mrm{u}} + \zeta T \right)Y_i \right)\,,
\end{equation}
\begin{equation}
	P\mrm{b} = h_\eta T^3\left( \frac{2}{3}\zeta T \sum_i Y_i \right)\,,
\end{equation}
where $\Delta M_i$ is the mass excess of nucleus $i$ (see Table \ref{table.nuclei} in Appendix \ref{appendix.nuclear}) and $\zeta = 3/2M\mrm{u}$. The parameter $h_\eta$ can be determined dynamically through $h_\eta\sim \rho\mrm{b}/T^3 \sim 1/a^3 T^3$ which implies the logarithmic relation:
\begin{equation}
	\frac{{d}\ln(h_\eta)}{{d}t} = -3\left( \frac{{d}\ln(a)}{{d}t} + \frac{{d}\ln(T)}{{d}t} \right)\,. \label{eq_h}
\end{equation}

\subsubsection{Nuclear reactions}

The set of nuclear reactions used in \texttt{AlterBBN} is given in Tables \ref{table.nuclear1}--\ref{table.nuclear3} in Appendix \ref{appendix.nuclear}. Each one of them can be written under the generalised form (to take into account reactions where 3 nuclei are involved \cite{ThesisEspen}):
\begin{equation}
	N_i\,\, ^{A_i}Z_i + N_j\,\, ^{A_j}Z_j + N_k\,\, ^{A_k}Z_k \leftrightarrow N_l\,\, ^{A_l}Z_l + N_m\,\, ^{A_m}Z_m + N_n\,\, ^{A_n} Z_n\,,
\end{equation}
where $N_i$ is the number of nuclei $Z_i$ that enters into the reaction and $A_i$ is their atomic number (see Table \ref{table.nuclei} in Appendix \ref{appendix.nuclear}). Then the abundance evolution of any nuclei $i$ is given by the equation:
\begin{equation}
	\frac{{d}Y_i}{{d}t} = N_i \sum_{j,k,l,m,n} \left( -\frac{Y_i^{N_i} Y_j^{N_j}Y_k^{N_k}}{N_i!N_j!N_k!}\Gamma_{ijk\rightarrow lmn} + \frac{Y_l^{N_l}Y_m^{N_m}Y_n^{N_n}}{N_l!N_m!N_n!}\Gamma_{lmn\rightarrow ijk} \right)\,, \label{eq_Y}
\end{equation}
where $\Gamma_{ijk\rightarrow lmn}$ and $\Gamma_{lmn\rightarrow ijk}$ are the forward and reverse reaction rates respectively.

\subsubsection{Initial conditions}

The dynamical variables of interest are $h(t)$, $\phi_e(t)$ and $Y_i(t)$, all functions of time, or equivalently of temperature. The initial temperature is denoted by $T\mrm{i}$. The initial condition for $h$ depends on the initial value of the baryon-to-photon ratio $\eta\mrm{i}$ which is obtained from entropy conservation:
\begin{equation}
	h_\eta(T\mrm{i}) = M\mrm{u} \frac{n_\gamma(T\mrm{i})}{T\mrm{i}^3} \eta_0\left( 1 + \frac{s_{e^\pm}(T\mrm{i})}{s_\gamma(T\mrm{i})} \right)\,, \label{eq_init_h}
\end{equation}
where $\eta\mrm{0}$ is the CMB baryon-to-photon ratio and for any species the entropy density reads:
\begin{equation}
	s\mrm{sp} = \frac{\rho\mrm{sp} + P\mrm{sp}}{T\mrm{sp}}\,.
\end{equation}
The initial condition for $\phi_e$ is:
\begin{equation}
	\phi_e(T\mrm{i}) \approx \frac{\pi^2}{2}\frac{h_\eta(T\mrm{i})Y\mrm{p}}{M\mrm{u}z\mrm{i}^3}\frac{1}{\displaystyle\sum_{n=1}^{\infty}(-1)^{n+1}nL(nz\mrm{i})}\,,
\end{equation}
where $Y\mrm{p}$ is the initial proton abundance and $z\mrm{i}$ is the initial adimensioned electron mass. The initial proton and neutron abundances $Y\mrm{p}$ and $Y\mrm{n}$ are given by the equilibrium of the reaction p $\leftrightarrow$ n:
\begin{equation}
	\begin{matrix}
		Y\mrm{p}(T\mrm{i}) = \dfrac{1}{1+e^{-q/T\mrm{i}}}\,, && && Y\mrm{n}(T\mrm{i}) = \dfrac{1}{1+e^{q/T\mrm{i}}}\,, \label{eq_prot_neut}
	\end{matrix}
\end{equation}
where $q = m\mrm{n} - m\mrm{p}$ is the nucleon mass difference. A similar equilibrium equation is applied to find the (small) initial deuterium abundance \cite{1967ApJ...148....3W}. All the other nucleus abundances are 0 because they have not started to form yet. The initial time can be found from the initial temperature through an ``infinite temperature'' approximation \cite{1967ApJ...148....3W}:
\begin{equation}
	t\mrm{i} = \frac{\sqrt{12\pi G\sigma}}{T\mrm{i}^2}\,, \label{eq_init_time}
\end{equation}
where $\sigma = \pi^2/60$ is the Stefan-Boltzmann constant.

The set of Eqs. (\ref{eq_a}), (\ref{eq_rho}), (\ref{eq_phie}), (\ref{eq_h}) and (\ref{eq_Y}) will be solved by different integration methods, as described in Appendix~\ref{sec.integration_methods}.

\subsection{Modified cosmologies}

In this section, we present the several modified scenarios, compared to the standard cosmology BBN, 
that are implemented in the \texttt{AlterBBN} program.

\subsubsection{Modified expansion rate}
\label{subsec.mod_expansion}

The addition of any new component in the early Universe, such as WIMPs, equivalent neutrinos or any kind of effective ``dark density,'' has an impact on the Friedmann equation (\ref{eq_a}). Indeed, the total density $\rho\mrm{tot}$ from previous section receives a ``dark'' contribution $\rho\mrm{D}$:
\begin{equation}
	\rho\mrm{tot}\rightarrow \rho\mrm{tot}+\rho\mrm{D}\,,
\end{equation}
\begin{equation}
	H^2 = \frac{8\pi G}{3}\left( \rho\mrm{tot} + \rho\mrm{D} \right)\,. \label{eq_hubble_modif}
\end{equation}
This modification affects the computation of the Hubble parameter, as shown by Eq. (\ref{eq_hubble_modif}). For the case of a ``dark density'', the following parametrisation has been used \cite{2008PhLB..669...46A}:
\begin{equation}
	\rho\mrm{D}(T) = \kappa_\rho\, \rho_\gamma(T_0)\left( \frac{T}{T_0} \right)^{n_\rho}\,,
\end{equation}
where $T_0 = 1$ MeV, chosen as a typical energy scale to fit the BBN initial conditions. Thus, $\kappa_\rho$ is the ratio of the dark density to the photon density at this energy scale. $n_\rho$ is the decrease exponent of this dark density (4 for radiation, 3 for matter, \textit{etc.}). A temperature cut can be added below which this density is strictly 0. Since the dark fluid properties are determined by the temperature $T$, it is assumed to be in interaction with the plasma and thus enters the set of interacting components of Eq.~(\ref{eq:dlna3_dt_simp}).

The pressure $P\mrm{D}$ associated to $\rho\mrm{D}$ is calculated automatically from a combination of the conservation equation and the entropy density evolution:
\begin{equation}
 P\mrm{D} = s\, \frac{{d}T}{{d}s}\, \left( \frac{{d}\rho\mrm{D}}{{d}T} - \rho\mrm{D} \frac{{d}s}{{d}T}\right)\,,
\end{equation}
where $s$ is the entropy density. The total radiation entropy $s\mrm{rad}$ is parametrised through the effective relativistic entropy degrees of freedom $h\mrm{eff}$ as:
\begin{equation}
 s\mrm{rad}(T) = h\mrm{eff}(T)\frac{2\pi^2}{45}T^3\,.
\end{equation}
The $h\mrm{eff}(T)$ are tabulated in the directory \texttt{sgStar\_heff}.

Another consequence of this modification is the computation of the initial time in the \texttt{AlterBBN} program, as one can see from Eq. (\ref{eq_init_time}). As long as the density is larger at a given time, the Hubble parameter is larger, and thus the time is smaller. For a fixed temperature, a correction has to be applied to the initial time $t\mrm{i}$ following:
\begin{equation}
	t\mrm{i,D} = \frac{H\mrm{SBBN}}{H\mrm{D}}t\mrm{i}\,,
\end{equation}
where $H\mrm{SBBN}$ is the standard Hubble rate given in Eq. (\ref{eq_a}) and $H\mrm{D}$ the modified one given by Eq. (\ref{eq_hubble_modif}).

\subsubsection{Additional neutrino species}
\label{subsec.add_neut}

The \texttt{AlterBBN} program leaves the possibility to modify the number of Standard Model neutrino species through the value of $N_\nu$ (see Eq. (\ref{eq_nu})). It is possible to add equivalent neutrinos to this number through a contribution $\Delta N_\nu$: $N\mrm{tot} = N_\nu + \Delta N_\nu$.

In addition, there could be a neutrino degeneracy between the 3 neutrino species of the Standard Model, which leads to non-zero adimensioned chemical potentials $\xi_{\nu_1}$, $\xi_{\nu_2}$ and $\xi_{\nu_3}$, which are not necessarily equal. This will have two consequences.

The first one is a modification of Eq.~(\ref{eq_nu}) for the computation of the neutrino density. The exact statistical mechanics formula is, for each species $i = 1,2,3$:
\begin{equation}
	\rho_{\nu_i,\overline{\nu}_i} = \frac{1}{2\pi^2}T_\nu^4 \int_0^{+\infty} {d}x \frac{x^3}{1+\exp(x\mp\xi_{\nu_i})}\,, \label{eq_full_neut}
\end{equation}
which for small degeneracies ($\xi_{\nu_i}<0.3$) is approximated by the expansion \cite{Kawano1992}:
\begin{equation}
	\rho_{\nu_i} + \rho_{\bar{\nu}_i} \approx \frac{\pi^2}{15}T_\nu^4\left( \frac{7}{8} + \frac{15}{4\pi^2}\xi_{\nu_i}^2 + \frac{15}{8\pi^4}\xi_{\nu_i}^4 + \mathcal{O}(\xi_{\nu_i}^4) \right)\,,
\end{equation}
and for high degeneracies ($\xi_{\nu_i}>30$) by the expansion \cite{Kawano1992}:
\begin{equation}
	\rho_{\nu_i} + \rho_{\bar{\nu}_i} \approx \frac{1}{8\pi^2}(T_\nu \xi_{\nu_i})^4 \left( 1 + \frac{2\pi^2}{\xi_{\nu_i}^2} + \mathcal{O}\left( \frac{1}{\xi_{\nu_i}^2} \right) \right)\,.
\end{equation}
For intermediate degeneracies ($0.3<\xi_{\nu_i}<30$), Eq. (\ref{eq_full_neut}) has to be integrated numerically.

The second consequence is a modification of the weak interaction equilibrium in the reaction p $\leftrightarrow$ n. The initial abundances of protons and neutrons, given in Eq. (\ref{eq_prot_neut}), become:
\begin{equation}
		Y\mrm{p}(T\mrm{i}) = \frac{1}{1+e^{-q/T\mrm{i}-\xi_{\bar\nu_e}}}, 
		\qquad
		Y\mrm{n}(T\mrm{i}) = \frac{1}{1+e^{q/T\mrm{i} + \xi_{\bar\nu_e}}}.
\end{equation}

\subsubsection{Modification of the entropy content}
\label{subsec.entropy_inj}

The early Universe content can also be modified by adding entropy components, which can derive from particle annihilation, or simply be any kind of ``dark entropy'' density. The relation between the scale factor and the temperature is then given by Eq.~(\ref{eq:dlna3_dt}), where $s$ is the total entropy density, which is composed of radiation and dark entropies, denoted by $s\mrm{rad}$ and $s\mrm{D}$, respectively.
Two different cases can occur (simultaneously or separately):

\paragraph{1 -- Dark entropy:}
the ``dark entropy'' $s\mrm{D}$ {is a function of the temperature $T$, but is not linked to any reheating process of radiation, so that the term $\partial s\mrm{D}/\partial t$ vanishes in Eq.~(\ref{eq:dlna3_dt}).} Two different parametrisations are possible. We can first parametrise it through \cite{2010JHEP...05..051A}:
\begin{equation}
	s\mrm{D} = \kappa_s s\mrm{\gamma}(T_0) \left( \frac{T}{T_0} \right)^{n_s}\,,
\end{equation}
where the energy scale at which the dark entropy to photon entropy ratio $\kappa_s$ is taken is also $T_0 = 1$ MeV, and $n_s$ is the decrease exponent of this component. A temperature cut below which this density is strictly 0 can be added. 

A second parametrisation of the ``dark entropy'' is possible through an additional entropy injection $\Sigma\mrm{D}$ \cite{2011JHEP...05..078A}:
\begin{equation}
	\Sigma\mrm{D}(T) = \kappa_\Sigma \Sigma\mrm{rad}^{\rm eff}(T_0)\left( \frac{T}{T_0} \right)^{n_\Sigma}\,,
\end{equation}
where $\kappa_\Sigma$ is the ratio of the dark entropy injection to the radiation entropy density time-derivative $\Sigma\mrm{rad}^{\rm eff}(T) \equiv \left| \dfrac{{d}s\mrm{rad}}{{d}t}\right| = 3 H s\mrm{rad}$ at $T_0 = 1$ MeV and $n_\Sigma$ is the decrease exponent of this production. The associated ``dark entropy'' density is thus the integral:
\begin{equation}
	s\mrm{D}(T) = 3\sqrt{\frac{5}{4\pi^3 G}}h\mrm{eff}(T)T^3 \displaystyle\int_0^T {d}T^\prime \frac{\sqrt{g_*(T)}\Sigma\mrm{D}(T^\prime)}{h\mrm{eff}^2(T^\prime)T^{\prime\,6}\sqrt{1+\rho\mrm{D}(T)/\rho\mrm{rad}(T)}}\,,
\end{equation}
where the values of $h\mrm{eff}(T)$ and $g\mrm{eff}(T)$ are tabulated in \texttt{sgStar\_heff}.

\paragraph{2 -- Reheating:}
the radiation entropy $s\mrm{rad}$ can receive contributions from a radiation entropy injection $\Sigma\mrm{rad}$ at constant temperature, such that $\partial s/\partial T = 0$ and
\begin{equation}
	\frac{{d}s\mrm{rad}}{{d}t} = -3 H s\mrm{rad} + \Sigma\mrm{rad}\,,
\end{equation}
which will modify the relation between the temperature and the time according to Eq.~(\ref{eq:dlna3_dt}), and result in a ``reheating'' of the primordial plasma and a local increase of the radiation entropy density.

We use the following parametrisation:
\begin{equation}
	\Sigma\mrm{rad}(T) = \kappa_{\Sigma\mrm{r}} \Sigma\mrm{rad}^{\rm eff}(T_0)\left( \frac{T}{T_0} \right)^{n_{\Sigma\mrm{r}}}\,,
\end{equation}
where $\kappa_{\Sigma_r}$ is ratio of the radiation entropy injection to the radiation entropy density time-derivative $\Sigma\mrm{rad}^{\rm eff}(T) \equiv \left| \dfrac{{d}s\mrm{rad}}{{d}t}\right| = 3 H s\mrm{rad}$ at $T_0 = 1$ MeV and $n_{\Sigma\mrm{r}}$ is the decrease exponent of this production. 

\subsubsection{Decaying scalar field}
\label{subsec.decay_scalar}

In this scenario, a primordial scalar field is decaying, as described in \cite{Arbey:2018uho}. Its density $\rho_\phi$ follows the Boltzmann equation:
\begin{equation}
 \frac{{d}\rho_\phi}{{d}t} = - n H \rho_\phi  - \Gamma_\phi \rho_\phi\,,
\end{equation}
where $\Gamma_\phi$ is the decay width of the scalar field and $n$ the decrease exponent of the scalar field density in term of the expansion factor. The scalar field decay results in radiation entropy injection at constant temperature such that $\partial s/\partial T = 0$ and:
\begin{equation}
\frac{{\partial}s\mrm{rad}}{{\partial}t} = -3 H s\mrm{rad} + \frac{\Gamma_\phi \rho_\phi}{T}\,.
\end{equation}
The decay width can be related to the reheating temperature $T\mrm{RH}$ through \cite{2006PhRvD..74b3510G}:
\begin{equation}
 \Gamma_\phi = \sqrt{\frac{4\pi^3 g\mrm{eff}(T\mrm{RH})}{45}} \, \frac{T^2\mrm{RH}}{M\mrm{P}}\,,
\end{equation}
where $g\mrm{eff}$ is the effective relativistic energy degrees of freedom, which is obtained from the tables contained in \texttt{sgStar\_heff}, and $M\mrm{P}$ is the Planck mass.

This scenario requires two input parameters, the first one being $\tilde{\rho}_\phi$ the scalar field energy density proportion to the photon energy density at the initial temperature ($\sim 2.3$ MeV) and the second one the reheating temperature $T\mrm{RH}$.

\subsubsection{WIMP scenarios}
\label{subsec.wimp}

Many WIMP scenarios have been implemented in \texttt{AlterBBN} \cite{ThesisEspen}. WIMPs are one of the candidates for the DM problem in cosmology. These are light, weakly interacting new particles characterised by their mass $m_\chi$, their type (Majorana or Dirac fermion, real or complex scalar) and their couplings to the SM (neutrinos, and possibly equivalent neutrinos, or EM interactions).

Many of the features and calculations needed to take WIMPs into account in \texttt{AlterBBN} are similar to those previously mentioned. It requires adding in a new WIMP density and pressure interacting with the plasma through self-annihilations. The WIMP density and pressure are given similarly to Eqs. (\ref{eq_rho_epm}) and (\ref{eq_P_epm}) by \cite{Chandra1939}:
\begin{equation}
	\rho_\chi = g_\chi m_\chi^4 \sum_{n=1}^{\infty} (-1)^{\beta(n+1)}\cosh(n\phi_\chi)M(nz_\chi)\,,
\end{equation}
\begin{equation}
	P_\chi = g_\chi m_\chi^4\sum_{n=1}^{\infty} \frac{(-1)^{\beta(n+1)}}{nz_\chi}\cosh(n\phi_\chi)L(nz_\chi)\,,
\end{equation}
where $g_\chi$ is the internal number of degrees of freedom of the WIMPs (1 for a real scalar, 2 for a complex scalar, 2 for a Majorana fermion, 4 for a Dirac fermion), $z_\chi = m_\chi/T_\chi$ is their adimensioned mass, and $\phi_\chi = \mu_\chi/T_\chi$ their adimensioned chemical potential. The temperature of the WIMPs $T_\chi$ is different ($T$ or $T_\nu$) depending on the SM couplings of the WIMP particles (EM or neutrinos, respectively). Also, $\beta = 0$ for bosonic WIMPs and $\beta = 1$ for fermionic WIMPs. Finally, the ``$\cosh$'' function has to be replaced by an ``$\exp$'' function in the case of self-conjugate particles (real scalars and Majorana fermions).

WIMPs also contribute to the entropy density of the early Universe and thus the initial condition for the $h_\eta(T\mrm{i})$ variable from Eq. (\ref{eq_init_h}) has to be modified:
\begin{equation}
	h_\eta(T\mrm{i}) = M\mrm{u}\frac{n_\gamma(T\mrm{i})}{T\mrm{i}^3} \eta_0\left( 1 + \frac{s_{e^\pm}(T\mrm{i}) + s_\chi(T\mrm{i})}{s_\gamma(T\mrm{i})} \right)\,,
\end{equation}

Finally, WIMPs may dynamically modify the neutrino temperature if they are coupled to them. For details on the way these modifications alter the differential equations of BBN, we refer the reader to Section 3.2 of Ref. \cite{ThesisEspen}.

\section{Content of the \texttt{AlterBBN} package}
\label{section.content}

The folder \texttt{alterbbn\_v2.X/} contains the 9 main programs of \texttt{AlterBBN}:
\begin{itemize}
	\item \texttt{stand\_cosmo.c},
	\item \texttt{alter\_eta.c},
	\item \texttt{alter\_neutrinos.c},
	\item \texttt{alter\_neutron.c},
	\item \texttt{alter\_etannutau.c},
	\item \texttt{alter\_standmod.c},
	\item \texttt{alter\_reheating.c},
	\item \texttt{alter\_phi.c},
	\item \texttt{alter\_wimps.c},
\end{itemize}
together with a \texttt{README} file, a \texttt{Makefile} file and a folder \texttt{alterbbn\_v2.X/src/}. The folder \texttt{alterbbn\_v2.X/src/} contains the source files:
\begin{itemize}
	\item \texttt{bbn.c},
	\item \texttt{bbnrate.c},
	\item \texttt{general.c},
	\item \texttt{cosmodel.c},
\end{itemize}
together with the files \texttt{include.h}, \texttt{numbers.h}, \texttt{bbn.h}, \texttt{bbnrate.h} -- containing the headers of all the program routines -- and \texttt{Makefile}. There is also a folder \texttt{alterBBN\_v2.X/src/sgStar\_heff/} containing numerically computed tables for $h\mrm{eff}(T)$ and $g\mrm{eff}(T)$, mentioned in Sections \ref{subsec.mod_expansion} and \ref{subsec.entropy_inj}, and a folder {\tt alterBBN\_v2.X/src/ contrib/newreac/} which contains routines to include reactions and isotopes from the REACLIB database \cite{2010ApJS..189..240C} into \texttt{AlterBBN}, as explained in Section \ref{subsec.reaclib}.

\subsection{Parameter structure}
\label{subsection.parameters}

There are two main parameter structures in the program \texttt{AlterBBN}, defined in the file \texttt{include.h}. The first one is:
\begin{verbatim}
typedef struct relicparam
/* structure containing the cosmological model parameters */
{
    int entropy_model,energy_model;
    double dd0,ndd,Tdend,Tddeq; // dark density
    double sd0,nsd,Tsend; // dark entropy
    double Sigmad0,nSigmad,TSigmadend; // dark entropy injection
    double Sigmarad0,nSigmarad,TSigmaradend; // standard entropy injection
    double nt0,nnt,Tnend; // non-thermal production of relics
    int coupd; // dark fluid coupling to plasma
    
    double quintn2,quintn3,quintn4,quintT12,quintT23,quintT34; 
    // effective quintessence model

    int phi_model; // decaying scalar field model switch
    double eta_phi,Gamma_phi,rhot_phi_Tmax,n_phi; // eta_phi = b / m_phi
    double rhot_phi0,Tphi0;
    double T_RH;
    double Sigmatildestar;
    double Sigmatildestar_max;
    double Tstdstar_max;

    double mgravitino; // gravitino mass
            
    double relicmass;
    int scalar;
    
    int solver; // switch for linear or logarithmic differential equation solver
    int beta_samples;
    
    double T; // Temperature in GeV
    double Y; // Y=n/s
    double Tfo,Tmax; // Freeze out and maximal temperature
    
    int full_comput; // Switch to deactivate the fast freeze out temperature
    determination
  
    double table_eff[276][3];   // Reads values from the SgStar files
   
    int use_table_rhoPD;
    double table_rhoPD[2][NTABMAX];
    int size_table_rhoPD;

    /*---------------------*/
    /* AlterBBN parameters */
    /*---------------------*/
    
    int err;
    int failsafe;               // Switch for the integration method
    double eta0;                // Initial Baryon to photon ratio
    double Nnu;                 // Number of Neutrinos (e+- included)
    double dNnu;                // Number of extra neutrinos (delta N_nu)
    double life_neutron,life_neutron_error;		// neutron lifetime
    double xinu1,xinu2,xinu3;	// [e-,neutrino], [muon,neutrino], 
    [tau,neutrino] respectively (degeneracy parameters)
    double m_chi;               // Mass of WIMP
    double g_chi;
    double Tinit;               // Initial temperature
    int wimp;                   // Switch to enable (1) / disable (0) wimps
    int SMC_wimp;               // wimp coupling to SM particles. 1 for EM, 
    2 for neutrino, 3 for neut. and eq. neut.
    int selfConjugate;          // 1/0 for self-conjugate/non-self-conjugate WIMP
    int fermion;
    int EM_coupled, neut_coupled, neuteq_coupled;
    double chi2;
    int nobs;    
    double fierz;		// Fierz interference term from LQ sector
    double B_chi;		// branching ratio of WIMP DM of mass m_p < m_chi < m_n 
    to explain the tau_n anomaly
    double rhob0;		// current baryon density
    double b_cdm_ratio;		// current ratio of baryon density to cold dark
    matter density 
}
relicparam; 
\end{verbatim}
and it contains all the parameters necessary to compute the BBN abundances of the elements, both in standard cosmology and in alternative cosmologies. This structure is common with \texttt{SuperIso Relic} \cite{2010CoPhC.181.1277A,2011CoPhC.182.1582A,2018arXiv180611489A} and some of its parameters are not used in \texttt{AlterBBN}.

The second one is:\newline
 \newline
\texttt{
typedef struct errorparam\newline
$\lbrace$\newline
	\indent int failsafe;\newline
	\indent int errnumber;\newline
	\indent double random[];\newline
	\indent double life\_neutron;\newline
$\rbrace$\newline
errorparam;
}\newline
 \newline
and it contains the parameters needed to give the estimated errors linked to the computed abundances.

\texttt{AlterBBN} has different modes to compute the abundance of the elements, determined by the \texttt{failsafe} variable of the \texttt{relicparam} structure: 0 corresponds to a fast but less precise calculation, and positive values to more precise but slower calculations. In case of a very non-standard cosmological scenario, it is advisable to set \texttt{failsafe} to 6 or more. By default, the standard mode is set to 1. A description of the different modes is provided in Section~\ref{sec.integration_methods}, together with the computation times and precision in Section~\ref{sec.integration_clocking}.

\subsection{Main routines}

The main routines defined in the library \texttt{libbbn.a} -- once compiled -- are listed below:
\begin{itemize}
	\item \texttt{void Init\_cosmomodel(struct relicparam* paramrelic)}\newline
	 \newline
	 This routine defined in \texttt{cosmodel.c} initialises the \texttt{paramrelic} structure with SBBN values. It sets the number of neutrino species to \texttt{Nnu} $= 3.046$ (including effects from non exactly relativistic $e^\pm$ \cite{2015PhRvD..91h3505N}), the baryon-to-photon ratio to \texttt{eta0} $= 6.09\times 10^{-10}$ \cite{PDG:2018}, the initial temperature to \texttt{Tinit} $= 27\times 10^9\,$K (corresponding to $2.3\,$MeV, an adequate value before the real start of BBN) and the lifetime of the neutron \texttt{life\_neutron} $= 880.2\,$s \cite{PDG:2018} (with its associated error to \texttt{life\_time\_error} $= 1.0\,$s). All the other parameters are set to 0.
	\item \texttt{void Init\_cosmomodel\_param(double eta, double Nnu,\newline double dNnu, double life\_neutron, double\newline life\_neutron\_error, double xinu1, double xinu2, double\newline xinu3, struct relicparam* paramrelic)}\newline
	 \newline
	 This routine defined in \texttt{cosmodel.c} specifies some parameters of the \texttt{paramrelic} structure with potentially non-standard values: the baryon-to-photon ratio \texttt{eta0}, the number of Standard Model neutrino species \texttt{Nnu}, the number of additional neutrino species \texttt{dNnu}, the eventual degeneracy of the Standard Model neutrinos \texttt{xinu1}, \texttt{xinu2} and \texttt{xinu3} (see Section \ref{subsec.add_neut}) and finally the neutron lifetime \texttt{life\_neutron} and the associated error \texttt{life\_neutron\_error}.
	\item \texttt{void Init\_dark\_density(double dd0, double ndd, double\newline T\_end, struct relicparam* paramrelic)}\newline
	 \newline
	 This routine defined in \texttt{cosmodel.c} specifies the parameters of the  \texttt{paramrelic} structure related to the effective dark density described in Section \ref{subsec.mod_expansion}. Here $\kappa_\rho =$ \texttt{dd0}, $n_\rho =$ \texttt{ndd} and \texttt{T\_end} is the temperature cutoff at which the effective dark density is set to 0.
	 \newline\newline
	\item \texttt{void Init\_dark\_entropy(double sd0, double nsd, double\newline T\_end, struct relicparam* paramrelic)}\newline
	 \newline
	 This routine defined in \texttt{cosmodel.c} specifies the parameters of the \texttt{paramrelic} structure related to the effective dark entropy density, in the case of no reheating, described in Section \ref{subsec.entropy_inj}. Here $\kappa_s =$ \texttt{sd0}, $n_s =$ \texttt{nsd} and \texttt{T\_end} is the temperature cutoff at which the effective dark entropy is set to 0.
	\item \texttt{void Init\_dark\_entropySigmaD(double Sigmad0, double\newline nSigmad, double T\_end, struct relicparam* paramrelic)}\newline
	 \newline
	 This routine defined in \texttt{cosmodel.c} specifies the parameters of the \texttt{paramrelic} structure related to the effective dark entropy production in the no-reheating case described in Section \ref{subsec.entropy_inj}. Here $\kappa_\Sigma =$ \texttt{Sigmad0}, $n_\Sigma =$ \texttt{nSigmad} and \texttt{T\_end} is the temperature cutoff at which the effective entropy production is set to 0.
	\item \texttt{void Init\_entropySigmarad(double Sigmarad0, double\newline nSigmarad, double T\_end, struct relicparam* paramrelic)}\newline
	 \newline
	 This routine defined in \texttt{cosmodel.c} specifies the parameters of the \texttt{paramrelic} structure related to the radiation entropy production in the reheating case, described in Section \ref{subsec.entropy_inj}. Here $\kappa_{\Sigma_r} =$ \texttt{Sigmarad0}, $n_{\Sigma_r} =$ \texttt{nSigmarad} and \texttt{T\_end} is the cutoff temperature at which the radiation entropy production is set to 0.
	\item \texttt{void Init\_scalarfield(double rhotilde\_phi, double T\_RH,\newline double eta\_phi, double n\_phi, struct relicparam* paramrelic)}\newline
	 \newline
	 This routine defined in \texttt{cosmodel.c} specifies the parameters of the \texttt{paramrelic} structure related to the decay of a scalar field during BBN, as described in Section \ref{subsec.decay_scalar}. Here \texttt{rhotilde\_phi} is the ratio of the scalar field density over the photon density at the initial temperature, \texttt{T\_RH} is the reheating temperature and \texttt{n\_phi} the decrease exponent of the scalar field density. The parameter \texttt{eta\_phi} has no effect in \texttt{AlterBBN} and is set to 0.
	\item \texttt{void Init\_wimp(double mass\_wimp, int EM\_coupled, int\newline neut\_coupled, int neuteq\_coupled, int fermion, int\newline selfConjugate, double g\_chi, struct relicparam* paramrelic)}\newline
	 \newline
	 This routine defined in \texttt{cosmodel.c} specifies the parameters of the \texttt{paramrelic} structure related to the existence of WIMPs during BBN, as described in Section \ref{subsec.wimp}. Here the parameters related to WIMP injection are specified: the WIMP mass $m_\chi =$ \texttt{mass\_wimp}, the SM couplings \texttt{EM\_coupled}, \texttt{neut\_coupled} and \texttt{neuteq\_coupled} (all switches between 0/1 for inactive/active), and the type of WIMP particle \texttt{fermion} and \texttt{selfConjugate} (both switches between 0/1 which represent the 4 types of wimps described in paragraph \ref{subsec.wimp}).
	\item \texttt{void rate\_weak(double f[], struct relicparam* paramrelic,\newline struct errorparam* paramerror)}\newline
	 \newline
	 This routine defined in \texttt{bbnrate.c} computes the forward reaction rates of the $\beta$-decays corresponding to the processes (2--11) given in Table \ref{table.nuclear1} in Appendix \ref{appendix.nuclear} and stores them into the (2--11) slots of the variable \texttt{f[]}. There is no reverse reaction so the slots (2--11) of the variable \texttt{r[]} do not need to be computed.
	\item \texttt{void rate\_pn(double f[], double r[], double T9, double Tnu, struct relicparam* paramrelic, struct errorparam* paramerror)}\newline
	 \newline
	 This routine defined in \texttt{bbnrate.c} computes the forward and reverse reaction rates of the nuclear reaction 1 (proton-neutron conversion) given in Table \ref{table.nuclear1} in Appendix \ref{appendix.nuclear} and stores them into the slot (1) of the variables \texttt{f[]} and \texttt{r[]}.
	\item \texttt{void rate\_all(double f[], double T9, struct relicparam*\newline paramrelic, struct errorparam* paramerror)}\newline
	 \newline
	 This routine defined in \texttt{bbnrate.c} computes the forward reaction rates of the nuclear reactions (12-100) given in Tables \ref{table.nuclear1} and \ref{table.nuclear2} in Appendix \ref{appendix.nuclear} and stores them into the (12-100) slots of the variable \texttt{f[]} (the reverse reaction rates will be estimated eslewhere through detailed balance factors contained in the variable \texttt{reacparam[][]} and stored in the variable \texttt{r[]}).
	\item \texttt{int nucl(struct relicparam* paramrelic, double ratioH[])}\newline
	 \newline
	 This routine defined in \texttt{bbn.c} is the main routine of the program, as it is the one that computes the BBN abundance ratios of all nuclei given in Table \ref{table.nuclei} in Appendix \ref{appendix.nuclear} and in particular the light elements \texttt{Yp} $= \rho(^4{\rm He})/\rho\mrm{b}$, \texttt{H2\_H} $= [^2{\rm H}]/[{\rm H}]$, \texttt{He3\_H} $= [^3{\rm He}]/[{\rm H}]$, \texttt{Li7\_H} $= [^7{\rm Li}]/[{\rm H}]$, \texttt{Li6\_H} $= [^6{\rm Li}]/[{\rm H}]$ and \texttt{Be7\_H} $= [^7{\rm Be}]/[{\rm H}]$ (note that \texttt{He3\_H} and \texttt{Li7\_H} contain the contributions of post-BBN decays of respectively \texttt{H3\_H} and \texttt{Be7\_H}). It returns 0 if the computation succeeded or 1 otherwise.
	\item \texttt{int bbn\_excluded(struct relicparam* paramrelic)}\newline
	 \newline
	 This routine defined in \texttt{bbn.c} is a ``container'' function that calls the \texttt{nucl} routine and compares its results with BBN observational constrains summarised in Appendix \ref{appendix.constraints}. It returns 0 if the constrains are satisfied, 1 if the abundances are not compatible with the observations and $-1$ if the computation fails.
	\item \texttt{int bbn\_excluded\_chi2(struct relicparam* paramrelic)}\newline
	 \newline
	 This routine defined in \texttt{bbn.c} is a ``container'' function that calls the \texttt{nucl} routine and performs a $\chi^2$ analysis of the $^2$H and $Y\mrm{p}$ values using their observational values and uncertainties given in Appendix \ref{appendix.chi2constraints}. It returns 0 if the confidence level of $95\,\%$ is satisfied, 1 if the abundances are not compatible with the observations and $-1$ if the computation fails.
\end{itemize}

\subsection{Error \& correlations}

\texttt{AlterBBN} includes the error estimation of the computations of the BBN abundances of the nuclei. This estimation relies on the parameters \texttt{err} and \texttt{life\_neutron\_error} contained in the \texttt{relicparam} structure and the parameters \texttt{errnumber} and \texttt{random[]} in the \texttt{errorparam} structure (see Section \ref{subsection.parameters}), as well as estimated errors on the nuclear reaction rates.

The \texttt{err} parameter switches between five methods of evaluation of the abundances of the elements and their errors:
\begin{itemize}
	\item \texttt{err} $= 0$: central values of the nuclear reaction rates are used for all reactions,
	\item \texttt{err} $= 1$: higher values are used for all reactions,
	\item \texttt{err} $= 2$: lower values are used for all reactions,
	\item \texttt{err} $= 3$: the covariance matrix is calculated through the variation of the parameters, using the higher value of the reaction rates and following the method of Ref.~\cite{2016JHEP...11..097A},
	\item \texttt{err} $= 4$: randomly Gaussian distributed values (between lower and higher) are used for all reactions.
\end{itemize}

The 8 main programs listed in the next section run successively the \texttt{err} $= 2$, \texttt{err} $= 0$ and \texttt{err} $= 1$ types in order to give associated ``lower", central and ``upper'' values of the abundances. Then they run the \texttt{err} $= 3$ type for all the reactions in order to compute a correlation matrix between the abundances, stored in the variable \texttt{corr\_ratioH[][]}.

An additional run of the \texttt{err} $= 4$ type is implemented to perform a Monte Carlo correlation analysis, but it is commented due to the extensive computational time (the abundance computation has to be done a lot of times, a number defined by the variable \texttt{niter} $= 1000$, by default).

The \texttt{err} $= 0$ type is used in order to compare the computed abundances to the observed ones (conservative limits). The \texttt{err} $= 3$ type is used to do a $\chi^2$ analysis (observational uncertainties).

\subsection{REACLIB reactions}
\label{subsec.reaclib}
By default, \texttt{AlterBBN} incorporates 26 elements and 100 nuclear reactions. A new module has been added in the latest version to easily include more elements and reactions from the JINA REACLIB database \cite{2010ApJS..189..240C}. The routines are contained in the directory \texttt{src/contrib/newreac/}. To generate a new set of isotopes and reactions, the program has to be compiled with \texttt{make}, which creates \texttt{create\_network.x}, and run with three parameters and one optional one: name of the WINVN file, name of the REACLIB database file, maximal atomic mass to be kept, and the optional parameter can be 0, 1 or 2, which selects all the isotopes with an atomic mass smaller than the maximal value, removes the isotopes with very small abundances after BBN, or removes the isotopes with very small abundances during BBN, respectively. The program generates three files, \texttt{numbers.h}, \texttt{bbn.h} and \texttt{bbnrate.h}, which contain the REACLIB parameters and are symbolically linked into \texttt{src/}. In the packages, the latest WINVN and READLIB files from \texttt{http://reaclib.jinaweb.org/} are included. Running\newline
\texttt{./create\_network.x winvn\_v2.0.dat results02200820.dat 30 1}\newline
adds 19 elements and 101 nuclear reactions. This set of new reactions is included by default in \texttt{AlterBBN}, but can be easily modified by running \texttt{create\_network.x}.

\section{Compilation and installation instructions}
\label{sec.install}

\texttt{AlterBBN} has been written in C respecting the \texttt{C99} standard, and it has been tested with the GNU C and the Intel C compilers on Linux, Windows and Mac. The package can be downloaded at the address:
\begin{center}
	\texttt{https://alterbbn.hepforge.org/}
\end{center}
The package should be unpacked in the desired directory, creating the main directory
\begin{center}
	\texttt{alterbbn\_vX.X/}
\end{center}
containing the material described in Section \ref{section.content}. If needed, the user's C compiler information and flags can be specified in the file \texttt{Makefile} in this main directory. In particular, some of the computations are made in parallel using the \texttt{OpenMP} library. The user should comment the corresponding lines if this library is not installed. More information is provided in the \texttt{README} file.

To activate the extra REACLIB reactions in \texttt{AlterBBN} (see Section \ref{subsec.reaclib}), the line \texttt{\#define REACLIB} has to be uncommented in \texttt{include.h}. If the number of extra elements is large, segmentation faults can occur because of the limited stack size. To circumvent the problem, it can be necessary to run \texttt{ulimit -s unlimited} before running the code.

To compile the library \texttt{libbbn.a}, type \texttt{make} in the main folder. The library file will be created in the subfolder \texttt{src/}. To compile a specific program, type \texttt{make name} or \texttt{make name.c} in the main folder, where \texttt{name} can be:
\begin{itemize}
	\item \texttt{stand\_cosmo} (see Section \ref{subsec.stand_cosmo}),
	\item \texttt{alter\_eta} (see Section \ref{subsec.alter_eta}),
	\item \texttt{alter\_neutrinos} (see Section \ref{subsec.alter_neutrinos}),
	\item \texttt{alter\_etannutau} (see Section \ref{subsec.alter_etannutau}),
	\item \texttt{alter\_standmod} (see Section \ref{subsec.alter_standmod}),
	\item \texttt{alter\_reheating} (see Section \ref{subsec.alter_reheating}),
	\item \texttt{alter\_phi} (see Section \ref{subsec.alter_phi}),
	\item \texttt{alter\_wimps} (see Section \ref{subsec.alter_wimp}).
	\item \texttt{alter\_neutron} (see Section \ref{subsec.alter_neutron}).
\end{itemize}

\section{Input and output description}

In this section we give input and output instructions for the 9 programs listed in Section \ref{sec.install}.

\subsection{Standard cosmology}
\label{subsec.stand_cosmo}

The program \texttt{stand\_cosmo.x} computes the BBN abundances of the nuclei as well as the associated errors and correlation matrix in the standard cosmological model. It takes one integer as argument; if it is 0 a fast calculation is done, larger values provide slower but more precise calculations. A description of the possible integration methods is given in Sections \ref{sec.integration_methods} and \ref{sec.integration_clocking}. The values of the baryon-to-photon ratio, the neutron lifetime and the number of neutrinos species are those fixed by the \texttt{Init\_cosmomodel} routine. Running the program with:\newline
\texttt{./stand\_cosmo.x 3}\newline
returns:\newline
\texttt{
\begin{tabular}{rrrrrrr}
			& Yp			& H2/H			& He3/H			& Li7/H			& Li6/H			& Be7/H 		\\
	 low:	& 2.473e-01     &  2.522e-05    &   1.024e-05   &    5.034e-10   &    1.685e-15   &    4.751e-10 	\\
	 cent:	& 2.472e-01     &  2.459e-05    &   1.033e-05   &    5.382e-10   &    1.083e-14   &    5.094e-10	\\
	 high:	& 2.471e-01     &  2.399e-05    &   1.044e-05   &    5.757e-10   &    3.512e-14   &    5.466e-10
\end{tabular}\newline
 \newline
--------------------\newline
With uncertainties:\newline
\begin{tabular}{rrrrrrr}
			& Yp			& H2/H			& He3/H			& Li7/H			& Li6/H			& Be7/H			\\
	 value:	& 2.472e-01    &   2.459e-05    &   1.033e-05    &   5.382e-10   &    1.083e-14   &    5.094e-10	\\
	 +/- :	& 3.201e-04    &   3.713e-07    &   1.648e-07    &   3.497e-11   &    1.083e-14   &    3.407e-11
\end{tabular}\newline
 \newline
Correlation matrix:\newline
\begin{tabular}{rrrrrrr}
			& Yp		& H2/H		& He3/H		& Li7/H		& Li6/H		& Be7/H		\\
	 Yp	 	& 1.000000     &   0.004790     &   0.026435    &    0.019322   &     0.001694   &     0.018095	\\
	 H2/H	& 0.004790    &    1.000000     &   -0.772793    &   -0.341440   &    0.064032    &    -0.354237	\\
	 He3/H	& 0.026435    &    -0.772793   &    1.000000     &   0.358671    &    -0.009010   &    0.370844	\\
	 Li7/H	& 0.019322    &    -0.341440   &    0.358671    &    1.000000    &    -0.024925   &    0.996514	\\
	 Li6/H	& 0.001694    &    0.064032    &    -0.009010   &    -0.024925   &    1.000000    &    -0.025749	\\
	 Be7/H	& 0.018095    &    -0.354237   &    0.370844    &    0.996514    &    -0.025749   &    1.000000
\end{tabular}\newline
 \newline
Compatible with BBN constraints (conservative limits)\newline
Compatible with BBN constraints (chi2 including correlations)
}

\subsection{Standard cosmology with modified parameters}

\subsubsection{Modification of the baryon-to-photon ratio}
\label{subsec.alter_eta}

The program \texttt{alter\_eta.x} computes the BBN abundances, errors and correlations in the standard cosmological model, but with a modified value of the baryon-to-photon ratio $\eta_0$, taken as an input argument. The next argument is optional and specifies the integration method. Running the program with:\newline
\texttt{./alter\_eta.x 3e-10}\newline
returns (hereafter, only the most relevant part of the output be given):\newline
\texttt{
\begin{tabular}{rrrrrrr}
			& Yp			& H2/H			& He3/H			& Li7/H			& Li6/H			& Be7/H			\\
	 value:	& 2.397e-01   &    7.545e-05    &   1.618e-05   &    1.318e-10    &   3.165e-14    &   6.869e-11	\\
	 +/- :	& 3.195e-04    &   1.289e-06    &   1.726e-07   &    1.100e-11    &   3.134e-14    &   5.272e-12	
\end{tabular}\newline
 \newline
Excluded by BBN constraints (conservative limits)\newline
Excluded by BBN constraints (chi2 including correlations)
}

\subsubsection{Modifications of the baryon-to-photon ratio, neutrino number and neutron lifetime}
\label{subsec.alter_etannutau}

The program \texttt{alter\_etannutau.x} computes the BBN abundances, errors and correlations in the standard cosmological model, but with a modified value of the baryon-to-photon ratio, number of Standard Model neutrino species, number of additional neutrino species, neutron lifetime, the optional integration method choice and an optional neutron lifetime error given in seconds, taken as input arguments in respective order. Running the program with:\newline
\texttt{./alter\_etannutau.x 6.09e-10 3.05 0.1 880.5 32 2.}\newline
returns:\newline
\texttt{
\begin{tabular}{rrrrrrr}
			& Yp			& H2/H			& He3/H			& Li7/H			& Li6/H			& Be7/H 		\\
	value:	& 2.488e-01    &   2.512e-05    &   1.041e-05   &    5.282e-10    &   1.115e-14   &    4.987e-10	\\
	+/- :	& 6.459e-03    &   5.415e-07    &   1.724e-07   &    3.549e-11    &   1.115e-14   &    3.430e-11
\end{tabular}\newline
 \newline
Compatible with BBN constraints (conservative limits)\newline
Compatible with BBN constraints (chi2 including correlations)
}

\subsubsection{Modifications of the neutrino number and degeneracies}
\label{subsec.alter_neutrinos}

The program \texttt{alter\_neutrinos.x} computes the BBN abundances, errors and correlations in the standard cosmological model, but with a modified number of Standard Model neutrino species, number of additional neutrino species and possibly three neutrino degeneracies $\xi_{\nu_i}$ (see Section \ref{subsec.add_neut}) as well as the optional integration method choice, taken as input arguments in respective order. Running the program with:\newline
\texttt{./alter\_neutrinos.x 3.046 0.1 0.1 0.1 0.1}\newline
returns:\newline
\texttt{
\begin{tabular}{rrrrrrr}
			& Yp			& H2/H			& He3/H			& Li7/H			& Li6/H			& Be7/H			\\
	value:	& 2.255e-01   &    2.354e-05    &   1.019e-05   &    5.046e-10   &    9.207e-15  &     4.799e-10	\\
	+/- :	& 2.965e-04   &    4.256e-07    &   1.678e-07   &    3.268e-11   &    9.181e-15  &     3.180e-11
\end{tabular}\newline
 \newline
Compatible with BBN constraints (conservative limits)\newline
Excluded by BBN constraints (chi2 including correlations)
}

\subsection{Modified expansion rate and entropy content}
\label{subsec.alter_standmod}

The program \texttt{alter\_standmod.x} computes the BBN abundances, errors and correlations in a cosmology scenario without reheating where the expansion rate and the entropy content are modified by the injection of a dark component throughout the BBN epoch (see Section \ref{subsec.entropy_inj}). It takes 4 to 8 input arguments ordered as $\kappa_\rho$, $n_\rho$, $\kappa_s$ and $n_s$, and possibly a switch to specify if the dark energy is coupled to the plasma, the cutoff temperatures in MeV for dark energy and dark entropy, and the integration method. Running the program with:\newline
\texttt{./alter\_standmod.x 0.1 3 0.1 4 0 0. 0.}\newline
returns:\newline
\texttt{
\begin{tabular}{rrrrrrr}
			& Yp			& H2/H			& He3/H			& Li7/H			& Li6/H			& Be7/H			\\
	value:	& 2.652e-01     &  4.249e-05     &  1.197e-05    &   3.244e-10   &    2.007e-14   &    2.754e-10	\\
	+/- :	& 3.243e-04     &  1.343e-06     &  1.978e-07    &   2.380e-11   &    1.996e-14   &    2.305e-11
\end{tabular}\newline
 \newline
Excluded by BBN constraints (conservative limits)\newline
Excluded by BBN constraints (chi2 including correlations)
}

\subsection{Effective reheating scenario}
\label{subsec.alter_reheating}

The program \texttt{alter\_reheating.x} computes the BBN abundances, errors and correlations in a cosmology scenario with reheating where dark energy and entropy are injected throughout the BBN epoch. It takes 5 to 10 input arguments ordered as $\kappa_\rho$, $n_\rho$, $\kappa_{\Sigma\mrm{r}}$, $n_{\Sigma\mrm{r}}$, the temperature cutoff for the injection in MeV and possibly $\kappa_s$, $n_s$, $\kappa_\Sigma$ and $n_\Sigma$ (see Section \ref{subsec.entropy_inj}), and the integration method. Running the program with:\newline
\texttt{./alter\_reheating 0 0 1 6 0.01}\newline
returns:\newline
\texttt{
\begin{tabular}{rrrrrrr}
	& Yp			& H2/H			& He3/H		& Li7/H 	& Li6/H			& Be7/H			\\
	value:	& 1.603e-01   &    3.469e-04   &    3.224e-05    &   2.222e-10   &    9.056e-14    &   1.662e-12	\\
	+/- :	& 2.730e-04   &    4.347e-06   &    2.825e-07    &   4.125e-11   &    8.706e-14    &   1.157e-13
\end{tabular}\newline
 \newline
Excluded by BBN constraints (conservative limits)\newline
Excluded by BBN constraints (chi2 including correlations)
}

\subsection{Decaying scalar field scenario}
\label{subsec.alter_phi}

The program \texttt{alter\_phi.x} computes the BBN abundances, errors and correlations in a cosmological scenario where a scalar field is decaying throughout the BBN epoch. It takes two input arguments which are the scalar field density ratio $\tilde{\rho}_\phi$ and the reheating temperature $T\mrm{RH}$ in MeV (see Section \ref{subsec.decay_scalar}), and two optional parameters, namely the scalar field decrease exponent (3 by default, matter-like behaviour) and the integration method. Running the program with:\newline
\texttt{./alter\_phi.x 0.1 10}\newline
returns:\newline
\texttt{
\begin{tabular}{rrrrrrr}
	& Yp			& H2/H			& He3/H		& Li7/H 	& Li6/H			& Be7/H			\\
	value:	& 2.454e-01    &   2.532e-05     &  1.040e-05    &   5.156e-10     &  1.104e-14   &    4.865e-10	\\
	+/- :	& 3.202e-04    &   1.040e-06     &  1.894e-07    &   4.650e-11     &  1.093e-14   &    4.596e-11 
\end{tabular}\newline
 \newline
Compatible with BBN constraints (conservative limits)\newline
Compatible with BBN constraints (chi2 including correlations)
}

\subsection{WIMP scenario}
\label{subsec.alter_wimp}

The program \texttt{alter\_wimps.x} computes the BBN abundances, errors and correlations in a cosmological scenario where WIMPs are added to the Standard Model particles. It takes 3 input arguments, which are the type of WIMP particle (1: real scalar, 2: complex scalar, 3: Majorana fermion, 4: Dirac fermion), the type of coupling to the Standard Model particles (1: neutrinos, 2: EM, 3: equivalent neutrinos) and $m_\chi$ (to be given in MeV) (see Section \ref{subsec.wimp}). An additional parameter can be given to specify the integration method. Running the program with:\newline
\texttt{./alter\_wimp.x 2 2 15.}\newline
returns:\newline
\texttt{
\begin{tabular}{rrrrrrr}
			& Yp			& H2/H			& He3/H			& Li7/H			& Li6/H			& Be7/H			\\
	value:	& 2.473e-01    &   2.421e-05    &   1.028e-05    &   5.496e-10   &    1.067e-14   &    5.212e-10	\\
	+/- :	& 3.183e-04    &   5.985e-07    &   1.712e-07    &   3.839e-11   &    1.065e-14   &    3.800e-11
\end{tabular}\newline
 \newline
Compatible with BBN constraints (conservative limits)\newline
Compatible with BBN constraints (chi2 including correlations)
}

\subsection{Neutron decay scenarios}
\label{subsec.alter_neutron}

The program \texttt{alter\_neutron.x} computes the BBN abundances, errors and correlations in a cosmological scenario where neutron beta decay is modified by beyond the Standard Model physics such as tensor or scalar currents, or dark decay channels with WIMPs near the neutron mass.

The reaction rates require phase space integrals with fermion state occupancy terms
	$1/(1 + e^{\pm x/z})$,
the electron occupancy factor,
where $x = E_e/m_e$ is the reduced electron energy, and $z = T_9 k_B / m_e$ is the dimensionless reduced final-state temperature, and
	$1/(1 + e^{\pm x_{\nu} / z_\nu \pm \xi_{\nu_e}})$,
the neutrino occupancy factor, 
where $x_\nu = q-x$ is the reduced neutrino energy,
and $z_\nu = T_\nu k_B / m_e$ is the dimensionless reduced neutrino temperature,
with $q = (m_n - m_p)/m_e \approx 2.53101$, the dimensionless neutron endpoint.
\texttt{alter\_neutron.x} increases the precision of the phase-space by adding the Fermi function that corrects for the electrostatic interaction of protons and betas:
\begin{equation}
	F(\pm\eta) = \frac{\pm\eta}{1 \pm e^{\pm\eta}}; \qquad \eta = 2\pi\alpha/\beta.
\end{equation}
It also allows for the careful calculation of scenarios with an added scalar or tensor interaction with the Fierz interference parameter 
for the free neutron, $b$ with the variable \texttt{fierz}, and a electron neutrino chemical potential $\xi_{\nu_e}$ using \texttt{xinu1},
as discussed in Section \ref{subsec.mod_expansion}.

The first of the four integrals calculated is just the standard neutron decay familiar from low energy,
\begin{equation}
	\Gamma_{n\to p e \bar\nu} 
		= \tilde{\Gamma}_0 \int_1^{\infty}
		    F(\eta) f(-x/z) f(-x_{\nu}/z_\nu - \xi_{\nu_e}) 
			\beta x_\nu^2 x (x+b) ~dx,
\end{equation}
where $f(x) = 1/(1+e^x)$ and $\tilde{\Gamma}_0$ is the phase space integral modified neutron decay rate at low temperature,
\begin{equation}
	(\tau_n \tilde{\Gamma}_0)^{-1} = \int_1^{q} F(\eta) \beta x_\nu^2 x (x+b) ~dx \approx 1.69174 +1.10855 ~ b,
\end{equation}

We then have the rates for the absorption of a positron or an electron neutrino
\begin{equation}
	\Gamma_{n \bar{e} \to p \nu} 
		= \tilde{\Gamma}_0 \int_1^{\infty}
		    f(x/z) f(-x_{\nu}/z_\nu - \xi_{\nu_e}) 
			\beta x_\nu^2 x (x-b) ~dx,
\end{equation}
and
\begin{equation}
	\Gamma_{n\nu \to pe} 
		= \tilde{\Gamma}_0 \int_1^{\infty} 
			F(-\eta) f(-x/z) f(x_{\nu}/z_\nu + \xi_{\nu_e}) 
			\beta x_\nu^2 x (x-b) ~dx,
\end{equation}
where $x_\nu=x+q$.
We also have the reverse reaction $p+e^-\to n+\nu$ which is \cite{Kolb:1990vq}
\begin{equation}
	\Gamma_{p e \to n \nu} 
		= \tilde{\Gamma}_0 \int_1^{\infty}
		    F(\eta) f(x/z) f(x_{\nu}/z_\nu + \xi_{\nu_e}) 
			\beta x_\nu^2 x (x+b) ~dx,
\end{equation}
where again as for the decay case, here $x_\nu=q-x$.
When \texttt{alter\_neutron.x} computes these integrals, it performs a numerical sum using the number of samples specified by \texttt{beta\_samples} variable. Hundreds to thousands of samples are needed to probe alterations to beta decay parameters.
\newline\texttt{./alter\_neutron.x 880.2 1.0}\newline
returns:\newline
\texttt{
\begin{tabular}{rrrrrrr}
			& Yp		& H2/H		& He3/H		& Li7/H		& Li6/H		& Be7/H		\\
	value:	& 2.483e-01     &  2.441e-05     &  1.032e-05     &  5.471e-10    &   1.081e-14       &  5.183e-10 \\
 	+/- :	& 3.185e-04     &  6.798e-07     &  1.816e-07     &  3.902e-11    &   1.079e-14       & 3.860e-11 \\
\end{tabular}\newline
 \newline
Compatible with BBN constraints (conservative limits)\newline
Compatible with BBN constraints (chi2 including correlations)
}

\section{Example of results}
\label{sec.result}

\begin{figure}[b!]
\centering
\includegraphics[width=8.05cm]{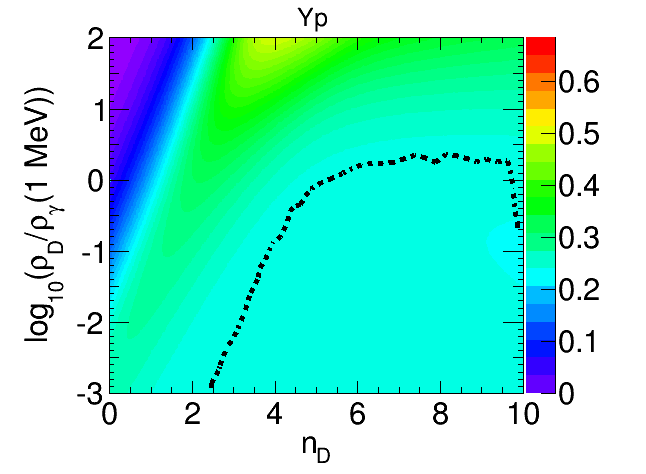}\includegraphics[width=8.05cm]{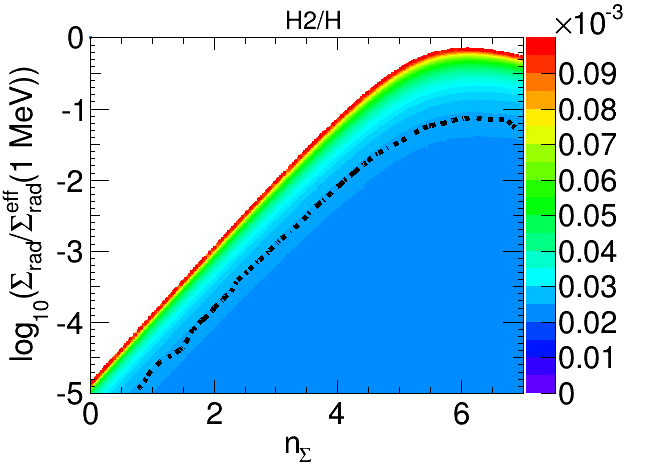}
\caption{$Y\mrm{p}$ in the dark density scenario (left) and [$^2$H]/[H] in the reheating scenario (right). The parameter regions excluded by BBN constraints are located above the black lines. The colours correspond to the values of $Y\mrm{p}$ and [$^2$H]/[H].\label{fig:BBN}}
\end{figure}%

To illustrate the capabilities of \texttt{AlterBBN}, we investigate the BBN constraints on the dark density and reheating scenarios, described in Sections \ref{subsec.mod_expansion} and \ref{subsec.alter_reheating}, respectively. For the first case, we can constrain the value of the the dark density at $T=1$ MeV and its exponent, and for the second case, we can constrain the value of the radiation injection at $T=1$ MeV and its exponent. We used \texttt{failsafe=3} for an improved precision. The results are shown in Fig.~\ref{fig:BBN}. The colour scale shows the value of $Y\mrm{p}$ in the dark density scenario, and of [$^2$H]/[H] in the reheating scenario. The black lines are the limits obtained using the constraints described in Appendix~\ref{appendix.chi2constraints}. The excluded region is above the lines.
 
\newpage

\appendix

\section{Nuclear reaction network}
\label{appendix.nuclear}

\begin{table}[h!]
	\footnotesize
	\centering{
		\caption{Table summarising the conventions used to denote the nuclei. Mass excess is given in MeV.\label{table.nuclei}}
		\begin{tabular}{|c|c|c|c|c|}
			\hline
			index 	& nuclei 	& atomic number 	& charge number 	& mass excess \\
			\hline
			1		& n			& 1					& 0					& $8.071388$ \\
			2		& p			& 1					& 1					& $7.289028$ \\
			3		& $^2$H		& 2					& 1					& $13.135825$ \\
			4		& $^3$H		& 3					& 1					& $14.949915$ \\
			5		& $^3$He	& 3					& 2					& $14.931325$ \\
			6		& $^4$He	& 4					& 2					& $2.424931$ \\
			7		& $^6$Li	& 6					& 3					& $14.0864$ \\
			8		& $^7$Li	& 7					& 3					& $14.9078$ \\
			9		& $^7$Be	& 7					& 4					& $15.7696$ \\
			10		& $^8$Li	& 8					& 3					& $20.9464$ \\
			11		& $^8$B		& 8					& 5					& $22.9212$ \\
			12		& $^9$Be	& 9					& 4					& $11.34758$ \\
			13		& $^{10}$B	& 10				& 5					& $12.05086$ \\
			14		& $^{11}$B	& 11				& 5					& $8.6680$ \\
			15		& $^{11}$C	& 11				& 6					& $10.6506$ \\
			16		& $^{12}$B	& 12				& 5					& $13.3690$ \\
			17		& $^{12}$C	& 12				& 6					& $0$ \\
			18		& $^{12}$N	& 12				& 7					& $17.3382$ \\
			19		& $^{13}$C	& 13				& 6					& $3.125036$ \\
			20		& $^{13}$N	& 13				& 7					& $5.3455$ \\
			21		& $^{14}$C	& 14				& 6					& $3.019916$ \\
			22		& $^{14}$N	& 14				& 7					& $2.863440$ \\
			23		& $^{14}$O	& 14				& 8					& $8.006521$ \\
			24		& $^{15}$N	& 15				& 7					& $0.101439$ \\
			25		& $^{15}$O	& 15				& 8					& $2.8554$ \\
			26		& $^{16}$O	& 16				& 8					& $-4.737036$ \\
			\hline
		\end{tabular}
	}
\end{table}

\begin{table}[p]
	\footnotesize
	\centering{
		\caption{Table summarising the conventions used to denote the nuclear reactions. Reverse reaction rates are detailed balance factors. Energy release is given in K.\label{table.nuclear1}}
		\begin{tabular}{|c|c|c|c|}
			\hline
			reaction index 	& reaction 												& reverse reaction rate 	& energy release 	\\
			\hline
			0 				& $n$ decay 													& $0$				& $0$ 				\\
			1 				& $n$ $\rightarrow$ p 											& $0$				& $0$ 				\\
			2 				& $^3$H $\rightarrow$ e$^-$ + $\overline{\nu}_e$ + $^3$He 		& $0$	 			& $0$				\\
			3 				& $^8$Li $\rightarrow$ e$^-$ + $\overline{\nu}_e$ + 2$^4$He 	& $0$ 				& $0$ 				\\
			4 				& $^{12}$B $\rightarrow$ e$^-$ + $\overline{\nu}_e$ + $^{12}$C 	& $0$ 				& $0$ 				\\
			5 				& $^{14}$C $\rightarrow$ e$^-$ + $\overline{\nu}_e$ + $^{14}$N 	& $0$ 				& $0$ 				\\
			6 				& $^{8}$B $\rightarrow$ e$^+$ + $\nu_e$ + 2$^{4}$He 			& $0$ 				& $0$ 				\\
			7 				& $^{11}$C $\rightarrow$ e$^+$ + $\nu_e$ + $^{11}$B 			& $0$ 				& $0$ 				\\
			8 				& $^{12}$N $\rightarrow$ e$^+$ + $\nu_e$ + $^{12}$C 			& $0$ 				& $0$ 				\\
			9 				& $^{13}$N $\rightarrow$ e$^+$ + $\nu_e$ + $^{13}$C 			& $0$ 				& $0$ 				\\
			10 				& $^{14}$O $\rightarrow$ e$^+$ + $\nu_e$ + $^{14}$N 			& $0$ 				& $0$ 				\\
			11 				& $^{15}$O $\rightarrow$ e$^+$ + $\nu_e$ + $^{15}$N 			& $0$ 				& $0$ 				\\
			12 				& H + n $\rightarrow$ + $^{2}$H 								& $0.477$ 			& $25.815$ 			\\
			13 				& $^{2}$H + n $\rightarrow$ $\gamma$ + $^{3}$H 					& $1.65$ 			& $72.612$ 			\\
			14 				& $^{3}$He + n $\rightarrow$ $\gamma$ + $^{4}$He 				& $2.63$ 			& $238.794$ 		\\
			15 				& $^{6}$Li + n $\rightarrow$ $\gamma$ + $^{7}$Li 				& $1.20$ 			& $84.132$ 			\\
			16 				& $^{3}$He + n $\rightarrow$ p + $^{3}$H 						& $1.001$ 			& $8.863$ 			\\
			17 				& $^{7}$Be + n $\rightarrow$ p + $^{7}$Li 						& $1.001$ 			& $19.080$ 			\\
			18 				& $^{6}$Li + n $\rightarrow$ $\alpha$ + $^{3}$H 				& $1.068$ 			& $55.503$ 			\\
			19 				& $^{7}$Be + n $\rightarrow$ $\alpha$ + $^{4}$He 				& $4.68$ 			& $220.382$ 		\\
			20 				& $^{2}$H + p $\rightarrow$ $\gamma$ + $^{3}$He 				& $1.65$ 			& $63.749$ 			\\
			21 				& $^{3}$H + p $\rightarrow$ $\gamma$ + $^{4}$He 				& $2.63$ 			& $229.931$ 		\\
			22 				& $^{6}$Li + p $\rightarrow$ $\gamma$ + $^{7}$Be 				& $1.20$ 			& $65.053$ 			\\
			23 				& $^{6}$Li + p $\rightarrow$ $\alpha$ + $^{3}$He 				& $1.067$ 			& $46.640$ 			\\
			24 				& $^{7}$Li + p $\rightarrow$ $\alpha$ + $^{4}$He 				& $4.68$ 			& $201.302$ 		\\
			25 				& $^{2}$H + $\alpha$ $\rightarrow$ $\gamma$ + $^{6}$Li 			& $1.55$ 			& $17.109$ 			\\
			26 				& $^{3}$H + $\alpha$ $\rightarrow$ $\gamma$ + $^{7}$Li 			& $1.13$ 			& $28.629$ 			\\
			27 				& $^{3}$He + $\alpha$ $\rightarrow$ $\gamma$ + $^{7}$Be 		& $1.13$ 			& $18.412$ 			\\
			28 				& $^{2}$H + d $\rightarrow$ n + $^{3}$He 						& $1.73$ 			& $37.934$ 			\\
			29 				& $^{2}$H + d $\rightarrow$ p + $^{3}$H 						& $1.73$ 			& $46.798$ 			\\
			30 				& $^{3}$H + d $\rightarrow$ n + $^{4}$He 						& $5.51$ 			& $204.116$ 		\\
			31 				& $^{3}$He + d $\rightarrow$ p + $^{4}$He 						& $5.51$ 			& $212.979$ 		\\
			32 				& $^{3}$He + $^{3}$He $\rightarrow$ 2p + $^{4}$He 				& $3.35$ 			& $149.229$ 		\\
			33 				& $^{7}$Li + d $\rightarrow$ n + $\alpha$ + $^{4}$He 			& $9.81$ 			& $175.487$ 		\\
			34 				& $^{7}$Be + d $\rightarrow$ p + $\alpha$ + $^{4}$He 			& $9.83$ 			& $194.566$ 		\\
			35 				& $^{3}$He + $^{3}$H $\rightarrow$ $\gamma$ + $^{6}$Li 			& $2.47$ 			& $183.290$ 		\\
			36 				& $^{6}$Li + d $\rightarrow$ n + $^{7}$Be 						& $2.52$ 			& $39.237$ 			\\
			37 				& $^{6}$Li + d $\rightarrow$ p + $^{7}$Li 						& $2.52$ 			& $58.317$ 			\\
			38 				& $^{3}$He + $^{3}$H $\rightarrow$ d + $^{4}$He 				& $1.59$ 			& $166.181$ 		\\
			39 				& $^{3}$H + $^{3}$H $\rightarrow$ 2n + $^{4}$He 				& $3.34$ 			& $131.503$ 		\\
			40 				& $^{3}$He + $^{3}$H $\rightarrow$ n + p + $^{4}$He 			& $3.34$ 			& $140.366$ 		\\
			\hline
		\end{tabular}
	}
\end{table}

\begin{table}[p]
	\footnotesize
	\centering{\caption{Second part of table \ref{table.nuclear1}.\label{table.nuclear2}}
		\begin{tabular}{|c|c|c|c|}
			\hline
			reaction index 	& reaction 													& reverse reaction rate & energy release	\\
			\hline
			41 				& $^{7}$Li + $^{3}$H $\rightarrow$ n + $^{9}$Be 				& $3.55$ 			& $121.136$ 		\\
			42 				& $^{7}$Be + $^{3}$H $\rightarrow$ p + $^{9}$Be 				& $3.55$ 			& $140.215$ 		\\
			43 				& $^{7}$Li + $^{3}$He $\rightarrow$ p + $^{9}$Be 				& $3.55$ 			& $129.999$ 		\\
			44 				& $^{7}$Li + n $\rightarrow$ $\gamma$ + $^{8}$Li 				& $1.33$ 			& $23.589$ 			\\
			45 				& $^{10}$B + n $\rightarrow$ $\gamma$ + $^{11}$B 				& $3.07$ 			& $132.920$ 		\\
			46 				& $^{11}$B + n $\rightarrow$ $\gamma$ + $^{12}$B 				& $2.37$ 			& $39.111$ 			\\
			47 				& $^{11}$C + n $\rightarrow$ p + $^{11}$B 						& $1.001$			& $32.086$ 			\\
			48 				& $^{10}$B + n $\rightarrow$ $\alpha$ + $^{7}$Li 				& $0.755$ 			& $32.371$ 			\\
			49 				& $^{7}$Be + p $\rightarrow$ $\gamma$ + $^{8}$B 				& $1.32$ 			& $1.595$ 			\\
			50 				& $^{9}$Be + p $\rightarrow$ $\gamma$ + $^{10}$B 			& $0.986$ 				& $76.424$ 			\\
			51 				& $^{10}$B + p $\rightarrow$ $\gamma$ + $^{11}$C 			& $3.07$ 				& $100.834$ 		\\
			52				& $^{11}$B + p $\rightarrow$ $\gamma$ + $^{12}$C 			& $7.10$ 				& $185.173$ 		\\
			53				& $^{11}$C + p $\rightarrow$ $\gamma$ + $^{12}$N 			& $2.37$ 				& $6.979$ 			\\
			54				& $^{12}$B + p $\rightarrow$ n + $^{12}$C 					& $3.00$ 				& $146.061$ 		\\
			55				& $^{9}$Be + p $\rightarrow$ $\alpha$ + $^{6}$Li 			& $0.618$				& $24.663$ 			\\
			56				& $^{10}$B + p $\rightarrow$ $\alpha$ + $^{7}$Be 			& $0.754$				& $13.291$ 			\\
			57				& $^{12}$B + p $\rightarrow$ $\alpha$ + $^{9}$Be 			& $0.291$ 				& $79.903$ 			\\
			58				& $^{6}$Li + $\alpha$ $\rightarrow$ $\gamma$ + $^{10}$B 	& $1.60$ 				& $51.761$ 			\\
			59				& $^{7}$Li + $\alpha$ $\rightarrow$ $\gamma$ + $^{11}$B 	& $4.07$ 				& $100.549$ 		\\
			60				& $^{7}$Be + $\alpha$ $\rightarrow$ $\gamma$ + $^{11}$C 	& $4.07$ 				& $87.543$ 			\\
			61				& $^{8}$B + $\alpha$ $\rightarrow$ p + $^{11}$C 			& $3.07$ 				& $85.948$ 			\\
			62				& $^{8}$Li + $\alpha$ $\rightarrow$ n + $^{11}$B 			& $3.07$ 				& $76.960$ 			\\
			63				& $^{9}$Be + $\alpha$ $\rightarrow$ n + $^{12}$C 			& $10.28$ 				& $66.158$ 			\\
			64				& $^{9}$Be + d $\rightarrow$ n + $^{10}$B 					& $2.06$ 				& $50.609$ 			\\
			65				& $^{10}$B + d $\rightarrow$ p + $^{11}$B 					& $6.42$ 				& $107.105$ 		\\
			66				& $^{11}$B + d $\rightarrow$ n + $^{12}$C 					& $14.85$ 				& $159.357$ 		\\
			67				& $^{4}$He + $\alpha$ + n $\rightarrow$ $\gamma$ + $^{9}$Be & $0.600$ 				& $18.262$ 			\\
			68				& $^{4}$He + 2a $\rightarrow$ $\gamma$ + $^{12}$C 			& $2.06$ 				& $84.420$ 			\\
			69				& $^{8}$Li + p $\rightarrow$ n + $\alpha$ + $^{4}$He 		& $3.54$ 				& $177.713$ 		\\
			70				& $^{8}$B + n $\rightarrow$ p + $\alpha$ + $^{4}$He 		& $3.55$ 				& $218.787$ 		\\
			71				& $^{9}$Be + p $\rightarrow$ d + $\alpha$ + $^{4}$He 		& $0.796$ 				& $7.554$ 			\\
			72				& $^{11}$B + p $\rightarrow$ 2a + $^{4}$He 					& $3.45$ 				& $100.753$ 		\\
			73				& $^{11}$C + n $\rightarrow$ 2a + $^{4}$He 					& $3.46$ 				& $132.838$ 		\\
			74				& $^{12}$C + n $\rightarrow$ $\gamma$ + $^{13}$C 			& $0.898$ 				& $57.400$ 			\\
			75				& $^{13}$C + n $\rightarrow$ $\gamma$ + $^{14}$C 			& $3.62$ 				& $94.884$ 			\\
			76				& $^{14}$N + n $\rightarrow$ $\gamma$ + $^{15}$N 			& $2.74$ 				& $125.715$ 		\\
			77				& $^{13}$N + n $\rightarrow$ p + $^{13}$C 					& $1.001$ 				& $34.846$ 			\\
			78				& $^{14}$N + n $\rightarrow$ p + $^{14}$C 					& $3.00$ 				& $7.263$ 			\\
			79				& $^{15}$O + n $\rightarrow$ p + $^{15}$N 					& $1.001$ 				& $41.037$ 			\\
			80				& $^{15}$O + n $\rightarrow$ $\alpha$ + $^{12}$C 			& $0.707$ 				& $98.659$			\\
			\hline
		\end{tabular}
	}
\end{table}

\begin{table}[p]
	\footnotesize
	\centering{\caption{Third part of table \ref{table.nuclear1}.\label{table.nuclear3}}
		\begin{tabular}{|c|c|c|c|}
			\hline
			reaction index 	& reaction 													& reverse reaction rate & energy release	\\
			\hline
			81				& $^{12}$C + p $\rightarrow$ $\gamma$ + $^{13}$N 			& $0.896$ 				& $22.554$ 			\\
			82				& $^{13}$C + p $\rightarrow$ $\gamma$ + $^{14}$N 			& $1.21$ 				& $87.621$ 			\\
			83				& $^{14}$C + p $\rightarrow$ $\gamma$ + $^{15}$N 			& $0.912$ 				& $118.452$ 		\\
			84				& $^{13}$N + p $\rightarrow$ $\gamma$ + $^{14}$O 			& $3.62$ 				& $53.705$ 			\\
			85				& $^{14}$N + p $\rightarrow$ $\gamma$ + $^{15}$O 			& $2.73$ 				& $84.678$ 			\\
			86				& $^{15}$N + p $\rightarrow$ $\gamma$ + $^{16}$O 			& $3.67$ 				& $140.733$ 		\\
			87				& $^{15}$N + p $\rightarrow$ $\alpha$ + $^{12}$C 			& $0.706$ 				& $57.622$ 			\\
			88				& $^{12}$C + $\alpha$ $\rightarrow$ $\gamma$ + $^{16}$O 	& $5.20$ 				& $83.111$ 			\\
			89				& $^{10}$B + $\alpha$ $\rightarrow$ p + $^{13}$C 			& $9.35$ 				& $47.134$ 			\\
			90				& $^{11}$B + $\alpha$ $\rightarrow$ p + $^{14}$C 			& $11.03$ 				& $9.098$ 			\\
			91				& $^{11}$C + $\alpha$ $\rightarrow$ p + $^{14}$N 			& $3.68$ 				& $33.921$ 			\\
			92				& $^{12}$N + $\alpha$ $\rightarrow$ p + $^{15}$O 			& $4.25$ 				& $111.620$ 		\\
			93				& $^{13}$N + $\alpha$ $\rightarrow$ p + $^{16}$O 			& $5.80$ 				& $60.557$ 			\\
			94				& $^{10}$B + $\alpha$ $\rightarrow$ n + $^{13}$N 			& $9.34$ 				& $12.288$ 			\\
			95				& $^{11}$B + $\alpha$ $\rightarrow$ n + $^{14}$N 			& $3.67$ 				& $1.835$ 			\\
			96				& $^{12}$B + $\alpha$ $\rightarrow$ n + $^{15}$N 			& $4.25$ 				& $88.439$ 			\\
			97				& $^{13}$C + $\alpha$ $\rightarrow$ n + $^{16}$O 			& $5.79$ 				& $25.711$ 			\\
			98				& $^{11}$B + d $\rightarrow$ p + $^{12}$B 					& $4.96$ 				& $13.296$ 			\\
			99				& $^{12}$C + d $\rightarrow$ p + $^{13}$C 					& $1.88$ 				& $31.585$ 			\\
			100				& $^{13}$C + d $\rightarrow$ p + $^{14}$C 					& $7.58$ 				& $69.069$ 			\\
			\hline
		\end{tabular}
	}
\end{table}

\newpage

\section{Description of the integration methods}
\label{sec.integration_methods}

The integration method can be modified using the {\tt failsafe} parameter of the {\tt relicparam} structure.

\subsection{Linearisation}

The abundance evolution of any nuclei $i$ is given by the Boltzmann equation:
\begin{equation}
	\frac{{d}Y_i}{{d}t} = N_i \sum_{j,k,l,m,n} \left( -\frac{Y_i^{N_i} Y_j^{N_j}Y_k^{N_k}}{N_i!N_j!N_k!}\Gamma_{ijk\rightarrow lmn} + \frac{Y_l^{N_l}Y_m^{N_m}Y_n^{N_n}}{N_l!N_m!N_n!}\Gamma_{lmn\rightarrow ijk} \right)\,.
\end{equation}

Unfortunately, the system for all the nuclei is composed of highly non-linear equations, and requires special attention. The system of equations has to be linearised, and written under the form:
\begin{equation}
\frac{d\mathbf{\vec{Y}}}{dt} = M(Y_i) \mathbf{\vec{Y}} \,, \label{eq:linearised}
\end{equation}
where $\mathbf{\vec{Y}}=(\mathbf{Y_1}, \cdots, \mathbf{Y_n})$ and $M$ is a matrix depending on the $Y_i$. The matrix elements can be obtained by a comparison with the form:
\begin{eqnarray}
\frac{{d}\mathbf{Y_i}}{{d}t} &=& N_i \sum_{j,k,l,m,n} \Bigl[ - \frac{1}{N_i! N_j! N_k! (N_i+N_j+N_k)} \Gamma_{ijk\rightarrow lmn} \\
&\times&\bigl[ N_i Y_i^{N_i-1} Y_j^{N_j} Y_k^{N_k} \mathbf{Y_i} + N_j Y_i^{N_i} Y_j^{N_j-1} Y_k^{N_k} \mathbf{Y_j} + N_k Y_i^{N_i} Y_j^{N_j} Y_k^{N_k-1}\mathbf{Y_k} \bigr]\nonumber\\
&+& \frac{1}{N_l! N_m! N_n! (N_l+N_m+N_n)}\Gamma_{lmn \rightarrow ijk}\nonumber\\
&\times&\bigl[ N_l Y_l^{N_l - 1} Y_m^{N_m} Y_n^{N_n} \mathbf{Y_l} + N_m Y_l^{N_l} Y_m^{N_m-1} Y_n^{N_n} \mathbf{Y_m} + N_n Y_l^{N_l} Y_m^{N_m} Y_n^{N_n-1}\mathbf{Y_n} \bigr]\Bigr] \,.\nonumber
\end{eqnarray}

\subsection{Stiff equation integration}

In a discrete integration, Eq.~(\ref{eq:linearised}) becomes
\begin{equation}
\frac{\mathbf{\vec{Y}_{n+1}}-\mathbf{\vec{Y}_n}}{\Delta t} = M(Y_n) \; \mathbf{\vec{Y}_n}
\end{equation}
where $n$ denotes the integration step number and $\Delta t$ the stepsize. This equation can be rewritten as:
\begin{equation}
\mathbf{\vec{Y}_{n+1}} = [1 + M(Y_n) \, \Delta t] \; \mathbf{\vec{Y}_n} \,.
\end{equation}
Unfortunately, $M$ has negative eigenvalues, and there is a high risk during the integration that the $(1 + M \, \Delta t)$ cancels. For this reason, we instead integrate as:
\begin{equation}
\mathbf{\vec{Y}_{n+1}} = [1 - M(Y_n) \, \Delta t]^{-1} \; \mathbf{\vec{Y}_n} \,,
\end{equation}
which is well-behaved and lead to a better convergence even for not too small values of $\Delta t$. The linearisation is therefore applied to the matrix $[1 - M(Y_n) \, \Delta t]$. To obtain the values of the derivatives of $Y_i$, $\mathbf{\vec{Y}_{n+1}}$ is obtain via a triangularisation of the matrix with a Cholesky decomposition, and inversion with a Gaussian elimination and back substitution. The derivatives of the abundances are finally given by $(\mathbf{\vec{Y}_{n+1}}-\mathbf{\vec{Y}_n})/\Delta t$.

\subsection{Runge-Kutta of order 2}
The methods 0--3 use a Runge-Kutta of order 2 integration such as:
\begin{equation}
\mathbf{\vec{Y}_{n+1}} = \mathbf{\vec{Y}_n} + \frac{d\mathbf{\vec{Y}}}{dt} \, \Delta t \,,
\end{equation}
where
\begin{equation}
\frac{d\mathbf{\vec{Y}}}{dt} = \frac12 \left[ \frac{d\mathbf{\vec{Y}}}{dt}(t_n) + \frac{d\mathbf{\vec{Y}}}{dt}(t_n+\Delta t) \right]\,.
\end{equation}
Following \cite{Kawano1992}, the stepsize is adapted with
\begin{equation}
 \Delta t = \min\left(\left|\frac{T}{dT/dt}\right| c_t, \left|\frac{Y_i}{dY_i/dt}\left[1+\left(\frac{\log(Y_i)}{\log(Y_{\rm min})}\right)\right]\right| c_y \right)\,,
\end{equation}
if $\Delta t>\Delta t_{\rm min}$, where $Y_{\rm min}=10^{-30}$, and $c_t$, $c_y$ and $\Delta t_{\rm min}$ are:
\begin{center}
$\begin{tabular}{|c|c|c|c|}
\hline
 Method (\texttt{failsafe}) & ~~~~~$c_t$~~~~~ & ~~~~~$c_y$~~~~~ & ~~$\Delta t_{\rm min}$(s)\\
  \hline
 0 (fastest) & 0.1 & 0.5 & $10^{-2}$\\
 1 (default) & 0.01 & 0.25 & $10^{-10}$\\
 2 & 0.005 & 0.1 & $10^{-10}$\\
 3 (slowest) & 0.001 & 0.05 & $10^{-10}$\\
 \hline
\end{tabular}$
\end{center}
These methods are the fastest ones, but lack a more robust convergence test in scenarios very far from the standard one, for which the other methods are preferred.

\subsection{Runge-Kutta of order 2 with half step test}
\label{subsec.rk2half}
The methods 5--7 use the Runge-Kutta of order 2 integration, but the convergence test is different: for each step, the variable are computed twice, once with a step of size $\Delta t$, and the second with two step sizes $\Delta t/2$. If the variable values differs by more than a tolerance {\texttt{prec}}, the stepsize is divided by 2 and the calculations starts over. Otherwise, the stepsize is obtained by:
\begin{equation}
 \Delta t \rightarrow 1.8 \times \min[1,\max(0.3,\texttt{minprec})] \times \Delta t\,,
\end{equation}
where
\begin{equation}
 \texttt{minprec} = \min\left( \left| \frac{\texttt{prec} \, \times \, \mathrm{variable\,(2\;steps)}}{\mathrm{variable\,(2\;steps)} - \mathrm{variable\,(1\;step)}} \right|\right) \;,
\end{equation}
where only the abundances larger than $Y_{\rm min}$ are considered in the test, in addition to the other physical variables.

The values of $Y_{\rm min}$ and the tolerance are, for the different methods:
\begin{center}
$\begin{tabular}{|c|c|c|}
\hline
 Method (\texttt{failsafe}) & ~~~~~~~~~~$Y_{\rm min}$~~~~~~~~~~ & Tolerance \texttt{prec}\\
  \hline
 5 (fastest) & $10^{-25}$ & 5\%\\
 6 (recommended) & $10^{-30}$ & 1\%\\
 7 (slowest) & $10^{-30}$ & 0.1\%\\
 \hline
\end{tabular}$
\end{center}

\subsection{Runge-Kutta of order 4 with half step test}
\label{subsec.rk4half}

The standard Runge-Kutta of order 4 corresponds, for one time step, to:
\begin{equation}
 y_{n+1} = y_n + \frac{\Delta t}{6} (k_1 + 2 k_2 +2 k_3 + k_4) \,,
\end{equation}
where
\begin{eqnarray}
 k_1 &=& \frac{dy}{dt}(t_n,y_n)\\
 k_2 &=& \frac{dy}{dt}\left(t_n+\frac{\Delta t}{2},y_n + \frac{\Delta t}{2}\, k_1\right)\\
 k_3 &=& \frac{dy}{dt}\left(t_n+\frac{\Delta t}{2},y_n + \frac{\Delta t}{2}\, k_2\right)\\
 k_4 &=& \frac{dy}{dt}(t_n+\Delta t,y_n + \Delta t \, k_1)\,.
 \end{eqnarray}
The stepsize is then adapted similarly to the method of Section~\ref{subsec.rk2half}. The values of $Y_{\rm min}$ and the tolerance are, for the different methods:
\begin{center}
$\begin{tabular}{|c|c|c|}
\hline
 Method (\texttt{failsafe}) & ~~~~~~~~~~$Y_{\rm min}$~~~~~~~~~~ & Tolerance\\
  \hline
 10 (fastest) & $10^{-25}$ & 5\%\\
 11 (recommended) & $10^{-30}$ & 1\%\\
 12 (slowest) & $10^{-30}$ & 0.1\%\\
 \hline
\end{tabular}$
\end{center}

\subsection{Runge-Kutta of order 4--5}

The explicit Runge-Kutta methods correspond, for one time step $\Delta t$, to:
\begin{equation}
 y_{n+1} = y_n + \Delta t \sum_{i=1}^s b_i k_i\,,
\end{equation}
where $s$ is the number of sub-steps and
\begin{eqnarray}
 k_1 &=& \frac{dy}{dt}(t_n,y_n)\\
 k_2 &=& \frac{dy}{dt}\left( t_n+c_2 \Delta t ,  y_n + \Delta t ( a_{21} k_1) \right)\nonumber\\
 k_3 &=& \frac{dy}{dt}\left( t_n+c_3 \Delta t , y_n + \Delta t ( a_{31} k_1 + a_{32} k_2) \right)\\
 & \vdots & \nonumber\\
 k_s &=& \frac{dy}{dt}\left( t_n+c_s \Delta t , y_n + \Delta t ( a_{s1} k_1 + a_{s2} k_2 + \cdots + a_{s,s-1} k_{s-1}) \right) \,.\nonumber
 \end{eqnarray}

The methods of order 4--5 consist in evaluating $y_{n+1}$ using simultaneously 4 and 5 sub-steps with common $c_i$ and $a_{ij}$, and use the two results to estimate the numerical error and adapt the stepsize.

In \texttt{AlterBBN}, if the difference is smaller than the tolerance \texttt{prec}, the stepsize is adjusted to
\begin{equation}
 \Delta t \rightarrow \min\Bigl(1.1,\max\bigl(2,0.84\,(\texttt{prec}\times\texttt{minprec})^{1/4}\bigr)\Bigl) \Delta t\,,
\end{equation}
otherwise:
\begin{equation}
 \Delta t \rightarrow \max\Bigl(0.9,\min\bigl(0.5,0.84\,(\texttt{prec}\times\texttt{minprec})^{1/4}\bigr)\Bigl) \Delta t\,,
\end{equation}
with
\begin{equation}
 \texttt{minprec} = \min\left( \left| \frac{\texttt{prec} \, \times \, \mathrm{variable\,(2\;steps)}}{\mathrm{variable\,(2\;steps)} - \mathrm{variable\,(1\;step)}} \right|\right) \;,
\end{equation}
where only the abundances larger than $Y_{\rm min}$ are considered in the test, in addition to the other physical variables.

\subsubsection{Fehlberg-Runge-Kutta method}

The parameters $c_i$ for this method are:
\begin{center}
$\begin{tabular}{|c||c|c|c|c|c|}
\hline
 ~~$i$~~ & 1 & 2 & 3 & 4 & 5\\
  \hline\hline
$c_i$ & 1/4 & 3/8 & 12/13 & ~~1~~ & 1/2\\
  \hline
\end{tabular}$
\end{center}
The $a_{ij}$ are:
\begin{center}
$\begin{tabular}{|c||c|c|c|c|c|}
\hline
 $a_{ij}$ & 1 & 2 & 3 & 4 & 5\\
  \hline\hline
 1 & 1/4 & & & & \\
  \hline
 2 & 3/32 & 9/32 & & & \\
  \hline
 3 & 1932/2197 & -7200/2197 & 7296/2197 & & \\
   \hline
 4 & 439/216 & -8 & 3680/513 & -845/4104 & \\
   \hline
 5 & -8/27 & 2 & -3544/2565 & 1859/4104 & -11/40  \\
   \hline
\end{tabular}$
\end{center}
The order 4 solution is computed with:
\begin{center}
$\begin{tabular}{|c||c|c|c|c|c|}
\hline
 ~~$i$~~ & 1 & 2 & 3 & 4 & 5\\
  \hline\hline
$b_i$ & 25/216 & 0 & 1408/2565 & 2197/4104 & -1/5\\
  \hline
\end{tabular}$
\end{center}
and the order 5 with:
\begin{center}
$\begin{tabular}{|c||c|c|c|c|c|c|}
\hline
 ~~$i$~~ & 1 & 2 & 3 & 4 & 5 & 6\\
  \hline\hline
$b_i$ & 16/135 & 0 & 6656/12825 & 28561/56430 & -9/50 & 2/55\\
  \hline
\end{tabular}$
\end{center}
The other \texttt{AlterBBN} parameters are, depending on \texttt{failsafe}:
\begin{center}
$\begin{tabular}{|c|c|c|}
\hline
 Method (\texttt{failsafe}) & ~~~~~~~~~~$Y_{\rm min}$~~~~~~~~~~ & Tolerance\\
  \hline
 20 (fastest) & $10^{-25}$ & 5\%\\
 21 (recommended) & $10^{-30}$ & 1\%\\
 22 (slowest) & $10^{-30}$ & 0.1\%\\
 \hline
\end{tabular}$
\end{center}

\subsubsection{Cash-Karp-Runge-Kutta method}

The parameters $c_i$ for this method are:
\begin{center}
$\begin{tabular}{|c||c|c|c|c|c|}
\hline
 ~~$i$~~ & 1 & 2 & 3 & 4 & 5\\
  \hline\hline
$c_i$ & 1/5 & 3/10 & 3/5 & ~~1~~ & 7/8\\
  \hline
\end{tabular}$
\end{center}
The $a_{ij}$ are:
\begin{center}
$\begin{tabular}{|c||c|c|c|c|c|}
\hline
 $a_{ij}$ & 1 & 2 & 3 & 4 & 5\\
  \hline\hline
 1 & 1/5 & & & & \\
  \hline
 2 & 3/40 & 9/40 & & & \\
  \hline
 3 & 3/10 & -9/10 & 6/5 & & \\
   \hline
 4 & -11/54 & 5/2 & -70/27 & 35/27 & \\
   \hline
 5 & 1631/55296 & 175/512 & 575/13824 & 44275/110592 & 253/4096 \\
   \hline
\end{tabular}$
\end{center}
The order 4 solution is computed with:
\begin{center}
$\begin{tabular}{|c||c|c|c|c|c|c|}
\hline
 ~~$i$~~ & 1 & 2 & 3 & 4 & 5 & 6\\
  \hline\hline
$b_i$ & 2825/27648 & 0 & 18575/48384 & 13525/55296 & 277/14336 & 1/4\\
  \hline
\end{tabular}$
\end{center}
and the order 5 with:
\begin{center}
$\begin{tabular}{|c||c|c|c|c|c|c|}
\hline
 ~~$i$~~ & 1 & 2 & 3 & 4 & 5 & 6\\
  \hline\hline
$b_i$ & 37/378 & 0 & 250/621 & 125/594 & 0 & 512/1771\\
  \hline
\end{tabular}$
\end{center}
The other \texttt{AlterBBN} parameters are, depending on \texttt{failsafe}:
\begin{center}
$\begin{tabular}{|c|c|c|}
\hline
 Method (\texttt{failsafe}) & ~~~~~~~~~~$Y_{\rm min}$~~~~~~~~~~ & Tolerance\\
  \hline
 30 (fastest) & $10^{-25}$ & $10^{-2}$\\
 31 (recommended) & $10^{-30}$ & $10^{-4}$\\
 32 (slowest) & $10^{-30}$ & $10^{-5}$\\
 \hline
\end{tabular}$
\end{center}

\section{Comparison of the integration methods}
\label{sec.integration_clocking}

In this section, we compare the different integration methods. The Runge-Kutta 4 method with $\mathtt{failsafe}=12$ is by construction the most precise (and slowest one), to which the results will be compared.

In the standard cosmological model, we compare the precision of one single BBN calculation:
\begin{center}
\hspace*{-1.cm}$\begin{tabular}{|c|c|c|c|c|c|c|}
\hline
\texttt{failsafe} & $Y_p (\times 10^{-1})$ & deviation & $\rm^2H/H (\times 10^{-5})$ & deviation & $^7\rm Li/H (\times 10^{-10})$ & deviation\\
\hline
\hline
0 & 2.462 & $-$0.010 & 2.343 & $-$0.120 & 5.680 & +0.310\\
\hline
1 & 2.472 & 0.000 & 2.431 & $-$0.032 & 5.473 & +0.103 \\
\hline
2 & 2.472 & 0.000 & 2.454 & $-$0.009 & 5.402 & +0.032 \\
\hline
3 & 2.472 & 0.000 & 2.459 & $-$0.004 & 5.382 & +0.012 \\
\hline
5 & 2.524 & +0.048 & 2.672 & +0.209 & 4.974 & $-$0.396\\
\hline
6 & 2.475 & +0.003 & 2.483 & +0.020 & 5.316 & $-$0.054 \\
\hline
7 & 2.472 & 0.000 & 2.462 & $-$0.001 & 5.372 & +0.002 \\
\hline
10 & 2.510 & +0.038 & 2.787 & +0.314 & 4.654 & $-$0.716 \\
\hline
11 & 2.476 & +0.004 & 2.488 & +0.025 & 5.304 & $-$0.066 \\
\hline
12 & 2.472 & ---    & 2.463 & ---    & 5.370 & --- \\
\hline
20 & 2.488 & +0.016 & 2.606 & +0.143 & 4.988 & $-$0.382 \\
\hline
21 & 2.479 & +0.007 & 2.521 & +0.058 & 5.212 & $-$0.158 \\
\hline
22 & 2.473 & +0.001 & 2.467 & +0.004 & 5.361 & $-$0.009 \\
\hline
30 & 2.489 & +0.017 & 2.619 & +0.146 & 4.969 & $-$0.401 \\
\hline
31 & 2.477 & +0.005 & 2.488 & +0.025 & 5.311 & $-$0.059 \\
\hline
32 & 2.473 & +0.001 & 2.470 & +0.007 & 5.351 & $-$0.019 \\
\hline
\end{tabular}$
\end{center}
The deviations refer to the differences between the values obtained with $\mathtt{failsafe}=12$. For comparison, the values and theoretical uncertainties calculated with $\mathtt{failsafe}=12$ are:
\begin{eqnarray}
 Y_p & = & (2.472 \pm 0.003) \times 10^{-1}\,,\\
 ^2H/H & = & (2.463 \pm 0.038) \times 10^{-5}\,,\\
 ^7Li/H & = & (5.370 \pm 0.352) \times 10^{-10}\,.\\
\end{eqnarray}

The computation times with different compilers (with \texttt{OpenMP} activated unless specified otherwise) on an Intel Core i7-6700HQ with 4 cores at 2.60GHz are for one BBN calculation are:
\begin{center}
$\begin{tabular}{|c||c|c|c|c|}
\hline
\texttt{failsafe} & gcc 8.3 & clang 7.0  & icc 19.0 & gcc w/o OpenMP\\
\hline
\hline
0 & 0.0292439 s & 0.0371051 s & 0.035774 s & 0.030885 s\\
\hline
1 & 0.0598671 s & 0.0702951 s & 0.04901 s & 0.0663319 s\\
\hline
2 & 0.12691 s & 0.14824 s & 0.105646 s & 0.139108 s\\
\hline
3 & 0.513578 s & 0.595697 s & 0.426564 s & 0.561526 s\\
\hline
5 & 0.241128 s & 0.279304 s & 0.201433 s & 0.260442 s\\
\hline
6 & 1.83213 s & 2.23646 s & 1.5879 s & 1.97117 s\\
\hline
7 & 16.9573 s & 20.3785 s & 14.0737 s & 17.8445 s\\
\hline
10 & 0.843263 s & 0.992097 s & 0.688307 s & 0.89699 s\\
\hline
11 & 7.81774 s & 9.06975 s & 6.40297 s & 8.35519 s\\
\hline
12 & 72.1262 s & 84.8538 s & 59.8514 s & 77.7767 s\\
\hline
20 & 0.157761 s & 0.176582 s & 0.133258 s & 0.169685 s\\
\hline
21 & 0.54873 s & 0.625423 s & 0.449118 s & 0.597708 s\\
\hline
22 & 4.84355 s & 5.62732 s & 4.11268 s & 5.26304 s\\
\hline
30 & 0.200886 s & 0.231516 s & 0.170542 s & 0.215176 s\\
\hline
31 & 1.53365 s & 1.79631 s & 1.29438 s & 1.68841 s\\
\hline
32 & 5.85279 s & 6.8085 s & 4.89614 s & 6.41176 s\\
\hline
\end{tabular}$
\end{center}
For a single BBN calculation, \texttt{OpenMP} does not decrease the computation time. However \texttt{OpenMP} decreases the computation times if REACLIB is activated and the number of elements is larger.\\
\\
The computation times of the correlation matrix are:
\begin{center}
$\begin{tabular}{|c||c|c|c|c|}
\hline
\texttt{failsafe} & gcc 8.3 & clang 7.0  & icc 19.0 & gcc w/o OpenMP\\
\hline
\hline
0 & 0.710522 s & 0.945136 s & 0.618748 s & 2.5015 s\\
\hline
1 & 1.98385 s & 2.63015 s & 1.73446 s & 7.17126 s\\
\hline
2 & 4.37474 s & 6.00581 s & 3.70563 s & 14.8925 s\\
\hline
3 & 17.5996 s & 22.5279 s & 14.915 s & 59.1254 s\\
\hline
5 & 8.0979 s & 9.99622 s & 6.89687 s & 27.2925 s\\
\hline
6 & 64.6826 s & 77.3424 s & 51.4998 s & 207.582 s\\
\hline
7 & 549.041 s & 721.656 s & 467.687 s & 1886.54 s\\
\hline
10 & 27.4793 s & 34.2517 s & 24.0487 s & 94.5529 s\\
\hline
11 & 256.467 s & 321.405 s & 218.789 s & 882.831 s\\
\hline
12 & 2344.66 s & 3094.74 s & 2062.58 s & 8312.86 s\\
\hline
20 & 4.97176 s & 7.24952 s & 4.54886 s & 17.6842 s\\
\hline
21 & 17.3305 s & 28.8574 s & 15.2907 s & 62.2697 s\\
\hline
22 & 151.593 s & 224.326 s & 135.464 s & 550.91 s\\
\hline
30 & 6.28095 s & 8.24639 s & 5.8179 s & 22.2994 s\\
\hline
31 & 48.7308 s & 67.6528 s & 44.2791 s & 176.329 s\\
\hline
32 & 188.842 s & 258.642 s & 175.523 s & 671.334 s\\
\hline
\end{tabular}$
\end{center}

\newpage

\section{BBN constraints}

\subsection{Conservative set}
\label{appendix.constraints}

A conservative set of constraints from Ref.~\cite{2006PhRvD..74j3509J} is defined in the function \texttt{bbn\_excluded}:
\begin{equation}
\begin{matrix}
0.240 < Y\mrm{p} < 0.258\;, && 1.2 \times 10^{-5} < [^2\rm H]/[H] < 5.3 \times 10^{-5}\;,\\ 0.57 < [^3\rm H]/[^2\rm H] < 1.52\;, &&  [^7\rm Li]/[H] > 0.85 \times 10^{-10}\;,\\ [^6\rm Li]/[^7\rm Li] < 0.66\;,
\end{matrix}
\end{equation}
which constrains the helium abundance $Y\mrm{p}$ and the primordial [$^2$H]/[H], [$^3$H]/[$^2$H], [$^7$Li]/[H] and [$^6$Li]/[$^7$Li] ratios.\\
The user can change these constraints in the routine \texttt{bbn\_excluded} which can be found in \texttt{src/bbn.c}.

\subsection{$\chi^2$ combination of recent constraints}
\label{appendix.chi2constraints}

The function \texttt{bbn\_excluded\_chi2} contains another recent set of uncorrelated observational measurements \cite{Aver:2015iza,Cooke:2017cwo,PDG:2018,Sbordone:2010zi}:

\begin{align}
Y\mrm{p} &= 0.2449 \pm 0.0040\,,\nonumber\\
[^2\rm H]/[H] &= (2.527\pm0.030) \times 10^{-5}\,,\\
[^3\rm H]/[^2\rm H] &= (1.1\pm 0.2) \times 10^{-5}\,,\nonumber\\
[^7\rm Li]/[H] &= (1.58 \pm 0.30) \times 10^{-10}\,.\nonumber
\end{align}

To assess the validity of the calculated abundances, a $\chi^2$ is computed using
\begin{equation}
 \chi^2 = \sum_i \, (O_i - E_i) \, C^{-1}_{ij} \, (O_j - E_j)\,,
\end{equation}
where $i$ corresponds to the abundances used to set the constraints, $O_i$ to the calculated abundance, $E_i$ to the observational measurement and $C_{ij}^{-1}$ to the inverse of the sum of the theoretical and experimental covariance matrices.

The number of degrees of freedom is considered to be the number of observational constraints, and the exclusion is assessed at 95\,\% C.L.

For the time-being, only the constraints of $Y\mrm{p}$ and $[^2\rm H]/[H]$ are used to compute the $\chi^2$. The user can change this in routine \texttt{bbn\_excluded\_chi2} which can be found in \texttt{src/bbn.c}.

\newpage

\bibliographystyle{JHEP}
\bibliography{biblio}

\end{document}